\documentclass[11pt]{article}
\usepackage{UF_FRED_paper_style}
\usepackage[toc,page]{appendix}
\usepackage[flushleft]{threeparttable}
\usepackage{amsmath}
\usepackage{amsfonts}
\usepackage{mathtools}
\usepackage{caption}

\newcommand{\ind}{\perp\!\!\!\!\perp}
\newcommand{\comment}[1]{}

\title{Who Benefits from Political Connections in Brazilian Municipalities%\thanks{I thank some people for comments.}
}

\author{\href{https://orcid.org/0000-0001-7168-2983}{\includegraphics[scale=0.06]{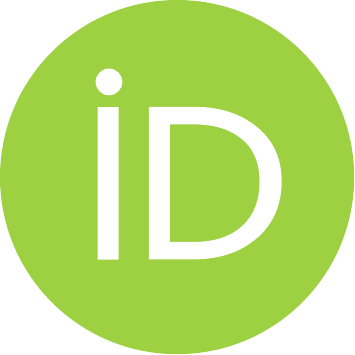}}\hspace{1mm}Pedro Forquesato\\% Name author
    \href{mailto:pforquesato@usp.br}{\texttt{pforquesato@usp.br}} \\
    School of Economics, Business and Accounting \\
    University of São Paulo %% Email author 
}
    
\date{\today}

\begin{document}

% %%%%%%%%%%%%%%%%%%%%%%%%%%%%%%%%%%%%%%%%%%%%%%%%%%%%%%%%%%
% %%%%%%%%%%%%%%%%%%%%%%%%%%%%%%%%%%%%%%%%%%%%%%%%%%%%%%%%%%
% ABSTRACT
% %%%%%%%%%%%%%%%%%%%%%%%%%%%%%%%%%%%%%%%%%%%%%%%%%%%%%%%%%%
% %%%%%%%%%%%%%%%%%%%%%%%%%%%%%%%%%%%%%%%%%%%%%%%%%%%%%%%%%%
{\setstretch{.8}
\maketitle
% %%%%%%%%%%%%%%%%%%
\begin{abstract}
% CONTENT OF ABS HERE--------------------------------------

A main issue in improving public sector efficiency is to understand to what extent public appointments are based on worker capability, instead of being used to reward political supporters (patronage). I contribute to a recent literature documenting patronage in public sector employment by establishing what type of workers benefit the most from political connections. Under the (empirically supported) assumption that in close elections the result of the election is as good as random, I estimate a causal forest to identify heterogeneity in the conditional average treatment effect of being affiliated to the party of the winning mayor. Contrary to previous literature, for most positions we find positive selection on education, but a negative selection on (estimated) ability. Overall, unemployed workers or low tenure employees that are newly affiliated to the winning candidate's party benefit the most from political connections, suggesting that those are used for patronage.

% END CONTENT ABS------------------------------------------
\noindent \vspace{5pt}

\end{abstract}
}

% %%%%%%%%%%%%%%%%%%%%%%%%%%%%%%%%%%%%%%%%%%%%%%%%%%%%%%%%%%
% %%%%%%%%%%%%%%%%%%%%%%%%%%%%%%%%%%%%%%%%%%%%%%%%%%%%%%%%%%
% BODY OF THE DOCUMENT
% %%%%%%%%%%%%%%%%%%%%%%%%%%%%%%%%%%%%%%%%%%%%%%%%%%%%%%%%%%
% %%%%%%%%%%%%%%%%%%%%%%%%%%%%%%%%%%%%%%%%%%%%%%%%%%%%%%%%%%

% --------------------
\section{Introduction}
% --------------------

Improving bureaucratic quality in developing countries is essential for better provision of public goods and for economic development \citep{finan2017personnel, rauch2000bureaucratic}. Worryingly, recent evidence has shown that political connections frequently determine entry into the government service, especially at the local level. As such, it becomes important to understand the effects of patronage on the quality of public sector employment, as it might significantly endanger state capacity and public policies.

Patronage in public hiring is when politicians use their appointment power of public workers to reward supporters or friends. There are many ways in which we can define political support: the literature has examined patronage with regards to party affiliation \citep{brollo2017victor, barbosa2019occupy, brassiolo2020my}, donations to the electoral process \citep{colonnelli2020patronage}, and to workers personally connected to the politician \citep{xu2018costs}. 

However, although the literature has established the importance of patronage for public employment policies, it gives little information on which individuals benefit from this patronage.\footnote{We discuss this point further at the end of this section.} This knowledge is important to understand the nature and consequences of politically motivated hiring. If workers hired for their political connection are nonetheless well educated and competent, then even if ubiquitous, patronage will not be very harmful to state capacity. Positive selection is possible if elections motivate politicians to show concern about government efficiency, or if good characteristics are correlated to the patronage mechanism (as in \cite{weaver2021jobs}). On the other hand, if politicians negatively select supporters for government employment, perhaps because a public sector job would be more valuable for supporters with worse private sector options, then we could expect patronage to be particularly damaging to public sector efficiency.

In this paper, I apply a causal forest machine learning technique \citep{athey2019generalized} to estimate the heterogeneous effect of being affiliated to the mayor's party on several measures of municipal employment. I estimate the conditional average treatment effect by taking advantage of the quasi-random nature of close elections. Namely, I assume that for sufficiently small intervals around the zero margin of victory, the election result is effectively random. Although I estimate causal forests using inverse probability weighting on observable variables, I show that prediction biases have a large mass on zero and propensity scores deviate little from a treatment probability of half. Both results give empirical support to our assumption of random assignment of close election results.

%% MAIN RESULTS

The main results are as follows. Firstly, while not the main purpose of our paper, I replicate and expand on previous findings by estimating that being affiliated to the party of the mayor causally increases the probability of municipal employment by 8.3 percentage points (on a baseline of 21\%), of which 4.1 p.p. are employment in managerial positions (baseline 1.6\%), 3.4 p.p. in white collar occupations (baseline 9.6\%) and 0.9 p.p. in blue collar occupations (baseline 9.7\%). Hence, political supporters are mostly employed in managerial positions (effectively tripling their probability of employment), which is consistent with mayors attempting to direct public policies to their ideologies. On the other hand, although there is a positive and significant effect on blue collar occupations, which are positions with smaller policy impact, thus more likely to be patronage appointments, it has a much smaller magnitude, being about 10\% over the baseline.

Our main purpose, however, is to investigate heterogeneous effects of political connections by worker characteristics. To accomplish this, I estimate best linear predictors of causal forest conditional average treatment (CATE) estimates. Namely, by each value leaf of worker characteristics, the causal forest estimates the causal effect of this type of worker being affiliated to the winning candidate's party, by comparing their outcomes to outcomes of workers on the same leaf that are affiliated to the losing candidate's party, always in close elections, finding the CATE for this particular group.

I begin by examining whether the conditional average treatment effect of being a political supporter of an elected mayor is higher for high-skill or low-skill workers. This comparison is important to understand the impact of political connections on public service quality. A highly positive differential impact for low-skill workers would mean that this type is unlikely to be hired to public service, except when they have good political connections. In that case, these connections would impact public bureaucracy negatively, worsening the quality of the government.

When looking at education, however, and unlike previous literature, I observe the opposite. Only for municipal blue collar occupations heterogeneous effects are higher for workers with incomplete high-school education. Note that these are low paying public jobs, unappealing to more highly educated employees. For all other municipal occupations, the effects of political connections are significantly higher for more educated workers, and this is especially the case for managerial public positions. Therefore, we rule out the case that low education workers can find municipal employment through political connections in positions where they would not be employed otherwise. In fact, the positive employment effect of political connections comes almost entirely from observably qualified candidates. 

Then, I examine the impact of previous employment characteristics on CATE. We find that political connections are more valuable for workers already working on the occupations they are eventually employed in the municipality. Again, this is consistent with these hires being not sorely to reward political supporters. \footnote{Part of this effect could be hysteresis of municipal employees already in these positions. (And indeed we find that political connections are more impactful for municipal employees.) But in Table \ref{table:private} I show that similar results are obtained when restricting the sample to workers previously in the private sector.} We also find that for all outcomes, treatment effects are smaller for high-tenure employees, who have more stable jobs, and thus less to gain from public appointments.

Although education is important, it is an incomplete measure of worker competence. Hence, I move further by measuring heterogeneity in the value of political connections from unobservable characteristics that affect wages. By using the large panel nature of our data, I estimate individual fixed-effects in a Mincerian equation and interpret them as measures of individual ability.\footnote{One could worry that these fixed-effects are correlated with our treatment. While I control for treatment status when estimating the Mincerian equation, I approach this concern more directly in Appendix Table \ref{table:ability_prev}, where I estimate worker ability only using pre-treatment observations. We obtain qualitatively similar results.} 

These measures are negatively related to conditional treatment effects, especially in blue collar occupations, indicating that political connections are more valuable for workers with lower wage-earning capacity. We caution, however, that private sector wages might be not perfectly correlated with characteristics desirable in public sector employment.

Importantly, in another new contribution, I find that mayors are significantly more likely to patron supporters recently affiliated to the party. This is evidence that these appointments are not due to supporters having the same ideology or the mayor bringing the party bureaucracy to the municipal government. More likely, this reflects mayors requesting personal supporters to affiliate to the party to assist them in party democracy and political campaigns, and later rewarding them with public positions (patronage). 

Overall, although the evidence points to patronage being a dominant motive of these appointments, my results are ambiguous on the impact of this patronage on bureaucratic efficiency. While political connection is more valuable for unemployed or informally employed and low ability workers, exactly those that have the most to gain from a municipal job, they are \textit{positively} selected in education and are more often selected from positions in the same occupation. These facts point towards mayors showing concern for worker efficiency in the municipal government, and arguably positively impacting public sector efficiency, even when favoring political supporters.

%% PREVIOUS LITERATURE 

Our paper relates to a growing literature on determinants of public bureaucracy efficiency, especially in developing countries \citep{moreira2021civil, akhtari2017political, dahis2020selecting, xu2018costs, weaver2021jobs, iyer2012traveling}. More closely, it adds to a growing literature causally estimating the effect of being a supporter of elected mayors on public employment in municipalities, with a focus on Brazil (\cite{brollo2017victor, colonnelli2020patronage, barbosa2019occupy, brassiolo2020my}.

Although papers in this literature often test heterogeneity regarding some selected worker characteristics, in those papers these characteristics are chosen arbitrarily, raising the concern of confirmation bias, while the linear projection of conditional treatment effects allows us to investigate heterogeneity reasonably free of researcher input. Also, previous research was often conflicting on the direction of this heterogeneity\footnote{For example, while \cite{colonnelli2020patronage} finds that patronage selects negatively on education, \cite{brollo2017victor} finds no differential effect. I actually find that for most positions the effect of political connections is stronger for more educated workers.}. Here, the use of machine learning allows us to investigate more avenues of heterogeneity then previous research, and do so jointly. While no heterogeneity analysis is causal, I improve significantly the reliability of this analysis by estimating best linear predictors of conditional treatment effects controlling for municipality and term fixed-effects and other sources of heterogeneity.

By using machine learning to estimate heterogeneous treatment effects, this paper also relates to a recent literature on machine learning tools for causal analysis, particularly estimating heterogeneity in treatment effects \citep{athey2019estimating, wager2018estimation, chernozhukov2018generic, davis2017using, davis2020rethinking, chernozhukov2018double, hsu2019testing, nie2021quasi, su2009subgroup}. Here, I apply these estimating procedures on quasi-experimental evidence, under the assumption that near the election threshold, assignment is random (see Section \ref{sec:estimation} for a discussion of the plausibility of this assumption).

% --------------------
\section{Institutional Setting}
% --------------------

I study patronage by elected politicians in Brazilian municipalities. This is a particularly interesting setting to analyze, because Brazil is a developing country with a comparatively insulated and professional civil service bureaucracy, making the existence of political influence in public sector appointments economically important, but also not obvious.

\paragraph{Brazilian municipalities}

Brazil is a federation democracy composed of a federal government, 26 states (and a federal district) and 5,570 municipalities. Cities are administered by a mayor and a city council, both elected (jointly) every four years. I use data for 2004, 2008, 2012, and 2016 municipal elections. Election in municipalities under 200,000 voters work in a plurality system, while cities above that threshold elect their mayors in a run-off majoritarian system. For simplicity of interpretation, I restrict attention to cities with single-turn elections. All our results, therefore, should be interpreted as pertaining to bureaucracies of small and medium-sized municipalities.

Unlike countries like United States, where the federal government is responsible for a large share of public expenditures, in Brazil most public goods, especially education and health care, are provided by municipalities. They account for almost 60\% of public employment, reaching in 2017 over 6.5 million jobs, or 6.25\% of Brazilian working force \citep{lopez2018atlas}. Although Brazil is frequently regarded as having a sizable public wage premium, it is mostly pertaining to federal and (to a lesser degree) state employees. In 2017, the average wage for municipality workers was 600 dollars per month, about twice the (then) Brazilian minimum wage.

\paragraph{Party affiliation and political parties}

Following \cite{brollo2017victor}, our measure of political support is affiliation to the mayor's party. I use that measure because it leads to a larger number of observations to train my causal random forest\footnote{As opposed to, for example, campaign donors and losing candidates, as in \cite{colonnelli2020patronage, brassiolo2020my}}. Indeed, party affiliation is pervasive in Brazil. In 2008, Brazil had almost 11 million party affiliates, over 6\% of the entire electorate. The three largest parties, PT, PSDB, and PMDB, had then almost 1.5 million affiliates each. Interestingly, new party affiliations are heavily concentrated on months exactly one year before the election. Presumably, this is due to mayoral candidates bringing supporters to influence intra-party democracy, and it is consistent with our findings that recently affiliated supporters are more likely to be rewarded with public jobs (see Section \ref{sec:results}).

% --------------------
\section{Data} \label{sec:data}
% --------------------

I analyze municipality employment using data from \emph{Relação Anual de Informações Sociais} (RAIS), an administrative data set gathered by the Brazilian federal government, which registers the universe of formal (public and private) labor relations in Brazil. Importantly, the data does not contain any information on informal employment relations. This will prevent us from being able to discern if workers not in the database are unemployed or working in (mostly low-paying) informal jobs. Fortunately, however, our main purpose is to compare municipal employment between affiliates of winning and losing parties in mayoral elections, and the data covers the universe of Brazilian public employees. I use data from 2003 to 2017.

I merge the (restricted) administrative data with open access party affiliation records by name, and I exploit the universal nature of our data (i.e., it is not a sample) by assigning every worker as unemployed (or informally employed) in the years when they do not appear at RAIS.

Then, I add electoral results data publicly available at the Brazilian Supreme Electoral Court (TSE) for the 2004, 2008, 2012, and 2016 mayoral elections for municipalities under 200,000 voters. I restrict our sample to affiliates to either the winning or second-place parties in each election, and I consider as outcome variables their employment characteristics exactly one year after the new mayor takes office. Here I take advantage that previous research has found that most of the patronage happens in the first year of the first term when the mayor takes office (see, e.g., \cite{brollo2017victor} or \cite{colonnelli2020patronage}). Based on the same evidence, I remove incumbents from my analysis.\footnote{Results including incumbents are available in Appendix Table \ref{tab:incumbent}, and they are quantitatively similar.} 

The benchmark analysis considers all workers, but I also create separate data sets for heterogeneous effects on workers previously employed in the (formal) private sector (Table \ref{table:private}). Moreover, according to the estimation strategy explained in Section \ref{sec:estimation}, I restrict the data to municipal elections where the margin of victory was under 5 percentage points.\footnote{Results for 2.5 p.p. and 1 p.p. thresholds are available in Appendix Tables \ref{table:main_025} and \ref{table:main_001}, respectively, and are quantitatively similar. Appendix Table \ref{table:main_nowork} and \ref{table:main_public} present results for previously unemployed and public sector workers, respectively.} I end up with 620,823 observations in the final data set.

Table \ref{tab:descriptive} presents descriptive statistics of the variables used in the benchmark analysis, as well as estimated outcomes $\hat{Y}$, propensity scores $\hat{W}$, individual treatment effects $\hat{\tau}$, and biases $\hat{b}$ of the generalized random forest estimation for the five outcomes on the full data (see Section \ref{sec:estimation} below for more details).

% Table created by stargazer v.5.2.2 by Marek Hlavac, Harvard University. E-mail: hlavac at fas.harvard.edu
% Date and time: qua, nov 17, 2021 - 15:28:50
\begin{table}[!h] \centering 
\begin{threeparttable}
\scriptsize
  \caption{Descriptive statistics for the full sample} 
  \label{tab:descriptive} 
  \small
\begin{tabular}{@{\extracolsep{5pt}}lccccccc} 
\\[-1.8ex]\hline 
\hline \\[-1.8ex] 
Statistic & \multicolumn{1}{c}{N} & \multicolumn{1}{c}{Mean} & \multicolumn{1}{c}{St. Dev.} & \multicolumn{1}{c}{Min} & \multicolumn{1}{c}{Pctl(25)} & \multicolumn{1}{c}{Pctl(75)} & \multicolumn{1}{c}{Max} \\ 
\hline \\[-1.8ex]
Municipal worker & 620,863 & 0.239 & 0.426 & 0 & 0 & 0 & 1 \\ 
Mun. blue collar & 620,863 & 0.096 & 0.295 & 0 & 0 & 0 & 1 \\ 
Wage (arsinh) & 620,863 & 1.933 & 2.283 & 0.000 & 0.000 & 4.267 & 9.469 \\ 
Mun. white collar & 620,863 & 0.107 & 0.309 & 0 & 0 & 0 & 1 \\ 
Mun. manager & 620,863 & 0.035 & 0.185 & 0 & 0 & 0 & 1 \\ 
Mayor supporter & 620,863 & 0.489 & 0.500 & 0 & 0 & 1 & 1 \\ 
Age & 620,863 & 41.275 & 11.347 & 18 & 32 & 50 & 65 \\ 
High-school inc. & 620,863 & 0.420 & 0.493 & 0 & 0 & 1 & 1 \\ 
University & 620,863 & 0.192 & 0.394 & 0 & 0 & 0 & 1 \\ 
Male & 620,863 & 0.654 & 0.476 & 0 & 0 & 1 & 1 \\
Ability & 620,863 & 0.079 & 0.701 & $-$3.954 & $-$0.404 & 0.455 & 5.086 \\ 
Years affiliated & 620,863 & 8.458 & 6.388 & 0 & 2 & 14 & 20 \\ 
Newly affiliated & 620,863 & 0.056 & 0.230 & 0 & 0 & 0 & 1 \\ 
Blue collar (lag) & 620,863 & 0.269 & 0.444 & 0 & 0 & 1 & 1 \\ 
Manager (lag) & 620,863 & 0.045 & 0.206 & 0 & 0 & 0 & 1 \\ 
Government job (lag) & 620,863 & 0.234 & 0.423 & 0 & 0 & 0 & 1 \\ 
Employed (lag) & 620,863 & 0.473 & 0.499 & 0 & 0 & 1 & 1 \\ 
Tenure (lag) & 620,863 & 3.474 & 6.351 & 0 & 0 & 4 & 48 \\ 
Estab. size (lag) & 620,863 & 2.797 & 3.500 & 0 & 0 & 6 & 10 \\ 
Term=2009 & 620,863 & 0.222 & 0.416 & 0 & 0 & 0 & 1 \\ 
Term=2013 & 620,863 & 0.283 & 0.450 & 0 & 0 & 1 & 1 \\ 
Term=2017 & 620,863 & 0.246 & 0.430 & 0 & 0 & 0 & 1 \\ 
Propensity score & 620,863 & 0.489 & 0.041 & 0.313 & 0.463 & 0.516 & 0.675 \\ 
Municipal empl. ($\hat{Y}$) & 620,863 & 0.241 & 0.285 & 0.006 & 0.042 & 0.281 & 0.967 \\ 
Municipal empl. ($\hat{\tau}$) & 620,863 & 0.087 & 0.076 & $-$0.073 & 0.033 & 0.118 & 0.490 \\ 
Municipal empl. ($\hat{b}$) & 620,863 & $-$0.003 & 0.012 & $-$0.102 & $-$0.006 & 0.002 & 0.096 \\ 
Wage ($\hat{Y}$) & 620,863 & 1.922 & 1.375 & 0.124 & 0.710 & 3.162 & 6.097 \\ 
Wage ($\hat{\tau}$) & 620,863 & 0.368 & 0.330 & $-$0.201 & 0.141 & 0.500 & 2.269 \\ 
Wage ($\hat{b}$) & 620,863 & $-$0.012 & 0.053 & $-$0.544 & $-$0.037 & 0.014 & 0.470 \\ 
Mun. blue collar ($\hat{Y}$) & 620,863 & 0.096 & 0.218 & 0.003 & 0.011 & 0.047 & 0.940 \\ 
Mun. blue collar ($\hat{\tau}$) & 620,863 & 0.010 & 0.011 & $-$0.075 & 0.004 & 0.015 & 0.116 \\ 
Mun. blue collar ($\hat{b}$) & 620,863 & $-$0.002 & 0.010 & $-$0.121 & $-$0.002 & 0.001 & 0.111 \\ 
Mun. white collar ($\hat{Y}$) & 620,863 & 0.108 & 0.198 & 0.003 & 0.015 & 0.078 & 0.902 \\ 
Mun. white collar ($\hat{\tau}$) & 620,863 & 0.035 & 0.035 & $-$0.086 & 0.013 & 0.048 & 0.305 \\ 
Mun. white collar ($\hat{b}$) & 620,863 & $-$0.002 & 0.009 & $-$0.111 & $-$0.002 & 0.001 & 0.098 \\ 
Mun. manager ($\hat{Y}$) & 620,863 & 0.036 & 0.054 & 0.002 & 0.012 & 0.038 & 0.568 \\ 
Mun. manager ($\hat{\tau}$) & 620,863 & 0.043 & 0.053 & $-$0.035 & 0.014 & 0.051 & 0.459 \\ 
Mun. manager ($\hat{b}$) & 620,863 & $-$0.0003 & 0.003 & $-$0.069 & $-$0.0004 & 0.0004 & 0.058 \\ 
\hline \\[-1.8ex] 
\end{tabular} 
\begin{tablenotes}
\item \emph{Notes:} Summary statistics for all variables used in the regression, except for affiliate party and municipality. Considers only municipal elections with vote margin within $\left[-5\%, +5\%\right]$. Complementary dummy variables (e.g. term=2005) are omitted, but can be inferred from the table. $\hat{Y}$ denotes predicted outcome variables, $\hat{\tau}$ are estimated individual treatment effects and $\hat{b}$ are estimated biases, and they are presented for all five main outcome variables. Descriptive statistics for all sub-samples is available upon request.
\end{tablenotes}
\end{threeparttable}
\end{table} 

As it is recognized to generally lead to better accuracy in machine learning methods, we standardize all features in our analysis. This leads to the convenient property that the intercept of the linear predictor of conditional treatment effects on our features represents the average treatment effect. However, it has the unfortunate impact of making our dummy coefficients harder to interpret. To account for this, we re-scale our coefficients (and standard errors) for all dummy variables to the 0-1 scale usual in economics.

Finally, to account for unobservable wage earning characteristics, we take advantage of the long panel nature of our data to estimate individual fixed effects in a polynomial Mincerian equation with full controls, and use these fixed effects as features in our random forest and linear projections. One potential concern here is of reverse causality, as our treatment could influence the estimated individual fixed effects. I deal with this concern in our benchmark analysis by controlling for treatment-period status, but in Appendix Table \ref{table:ability_prev} I show that results using only pre-treatment observations are qualitatively similar, and indeed even stronger.\footnote{I do not use this specification as benchmark because using only pre-treatment observations biases the sample to later years, with potentially non-obvious effects.}

% --------------------
\section{Estimation} \label{sec:estimation}
% --------------------

Consider an independent and identically distributed sample of size $n$, that contains pre-treatment covariates (features) $X_i$ of dimension $d$, a real-valued response $Y_i$ and a treatment $W_i \in \{0,1\}$. I employ the potential outcomes framework, and we are interested in estimating conditional average treatment effects (CATE) of the form $$\tau (x) \equiv \mathbb{E} \left[ Y_i(1) - Y_i(0) \ \left| \right. X_i = x \right],$$ where $Y_i (1)$ is the outcome $i$ gets when treated and $Y_i(0)$ when untreated.

The usual identification problem comes from the fact that we never observe both outcomes for the same individual, and therefore to identify $\tau(x)$ we need to make assumptions about the form of sample selection we observe. In this paper, I adopt the local randomization framework for regression in discontinuity analysis and assume that for a sample of close elections, treatment assignment (election victory) is effectively random. This assumption implies unconfoundedness: $\left\{ Y_i(1), \ Y_i(0) \right\} \ind W_i | X_i$, an assumption sufficient for identification of conditional average treatment effects in our environment.

I exploit the discontinuity in treatment assignment at zero margin of victory to estimate the causal effect of being affiliated to the party of the winning mayoral candidate, by comparing wages and the probability of working in municipal government of affiliates to parties where the candidate narrowly won the election to affiliates to parties where the candidate narrowly lost. Furthermore, since our causal forest estimation accounts for municipality fixed-effects, we estimate causal effects within each municipality. Our identification assumption is that electoral performance varies continuously with non-observable attributes, and therefore in a sufficiently close interval from zero margin of victory, our treatment $W_i$ (the mayor being affiliated to the same party as the worker) is as good as randomly assigned.

It is important to note that although average treatment effects are causally estimated by the quasi-experimental regression in discontinuity design framework I employ, the same is not true for heterogeneity. Nevertheless, our best linear predictor of CATE estimates allows us to control for municipality and term fixed-effects, party dummies, as well as estimate heterogeneity jointly.\footnote{To reduce the size of the features matrix, I perform mean encoding on the municipality fixed-effects \citep{johannemann2019sufficient}.} This is a significant improvement over previous literature that separately estimates heterogeneity on many dimensions (covariates) that are highly correlated among themselves (e.g., education and occupation), and therefore act as omitted variables to each other.

I consider separately five outcomes; namely, (i) being employed in the the municipal public sector; (ii) inverse hyperbolic sine of wages; (iii) being employed in a blue collar municipal occupation; (iv) being employed in a white collar municipal occupation; and (v) being employed in a managerial municipal occupation. For each of them, I train a generalized random forest algorithm with honest sample partition \citep{athey2016recursive}, using the package \verb+grf+ in \verb+R+ language  \citep{athey2019generalized}.\footnote{Each of our generalized random forest regressions takes on average about 18 hours to run in a 20 CPUs (cloud) computing cluster.} Generalized random forests use the idea of greedy recursive partitioning and sub-sample aggregation of random forests to locally estimate (in our application) heterogeneous treatment effects. See, for example, \citep{friedman2017elements, biau2016random} for introductions to random forests. 

Formally, if we observe the aforementioned i.i.d. samples $(Y_i, W_i, X_i)_{i=1}^{N}$, we are interested in estimating conditional average effects of the form $\beta (x) \equiv \mathbb{E} \left[ b_i | X_i = x \right]$, where $b_i$ are the heterogeneous random effects in the treatment equation $Y_i = W_i \cdot b_i + \varepsilon_i$. In our empirical exercise, the treatment $W_i$ is being affiliated to the same party as the mayor, covariates $X_i$ are pre-treatment socioeconomic characteristics of the worker, and outcomes $Y_i$ are one of five already mentioned post-election employment outcomes.

I nonparametrically estimate conditional expectations $Y(x) = \mathbb{E} \left[ Y_i | X_i = x \right]$ and $W(x) = \mathbb{E} \left[ W_i | X_i = x \right]$ using a boosting regression forest, and center the outcome and treatment by the keep-one-out expectations $\widetilde{W}_i = W_i - W^{(-1)}(X_i)$ and $\widetilde{Y}_i = Y_i - Y^{(-1)}(X_i)$. I then compute random forest weights for each value of $x$ and use them to do a projection of locally weighted outcomes on locally weighted treatment value \citep{athey2019generalized}. All sub-sampling is done clustered within municipality, as well as all the estimation of standard errors. To obtain intelligible results of heterogeneity, I present results for the best linear predictor of CATE. I also investigate the mean characteristics of the population quartile with largest and smallest conditional treatment effects in a classification analysis \citep{chernozhukov2018generic}.

Importantly, causal forests are estimated using honest partition \citep{athey2016recursive}, where one sub-sample is used to construct the leaves of the forest, and another different sub-sample is used to estimate treatment effects within each leaf. In this sense, a fortunate property of our setting is that the large number of observations allows sample partitioning with small increases in variance.

As mentioned above, for robustness, I estimate by boosting the propensity scores for treatment probability, and use them to correct our CATE estimates. Encouragingly, as I show in Figure \ref{fig:propensity} below, the estimated propensity scores are concentrated around half, and vary little with features, corroborating our identification strategy that treatment assignment is quasi-random. Moreover, in panels (b) and (c) of Figure \ref{fig:propensity} I report the same statistic for smaller vote margin intervals, where the identification strategy should be more credible, and find mostly indistinguishable results. 

\begin{figure}[h]
\begin{subfigure}{.3\textwidth}
    \centering
        \includegraphics[scale=0.28]{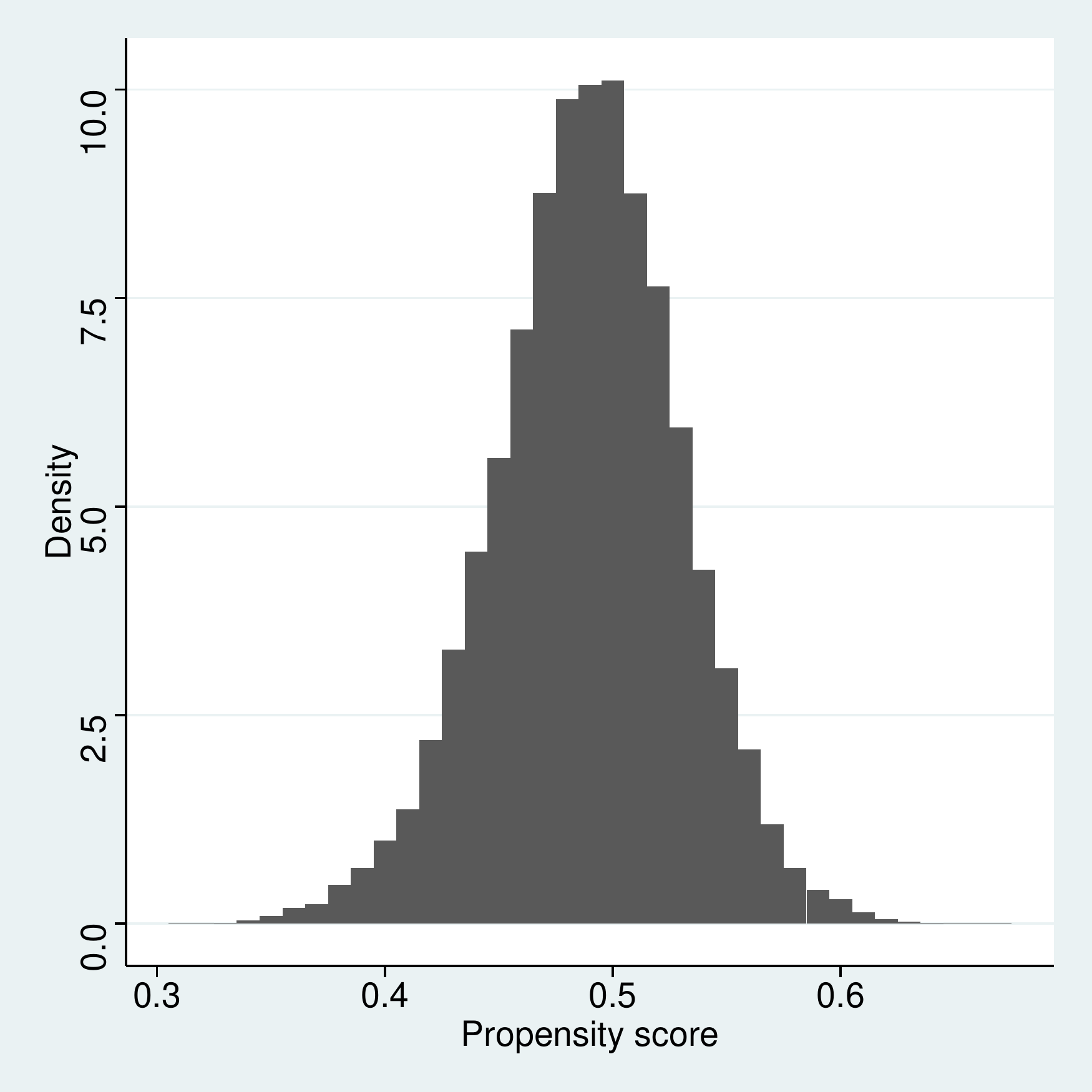}
    \caption{Benchmark}
\end{subfigure}
\begin{subfigure}{.3\textwidth}
    \centering
        \includegraphics[scale=0.28]{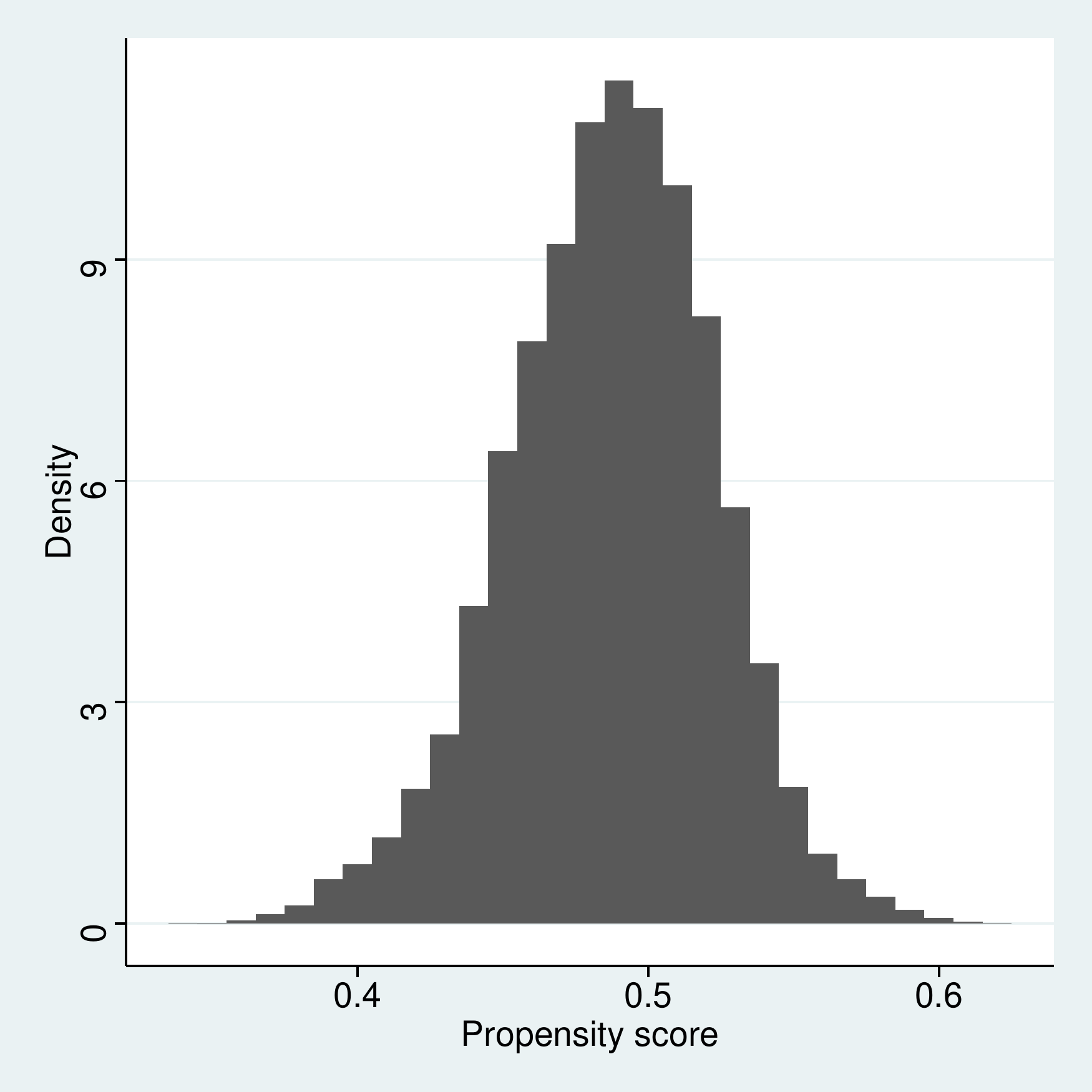}
    \caption{$\pm$ 2.5 p.p.}
\end{subfigure}
\begin{subfigure}{.3\textwidth}
    \centering
        \includegraphics[scale=0.28]{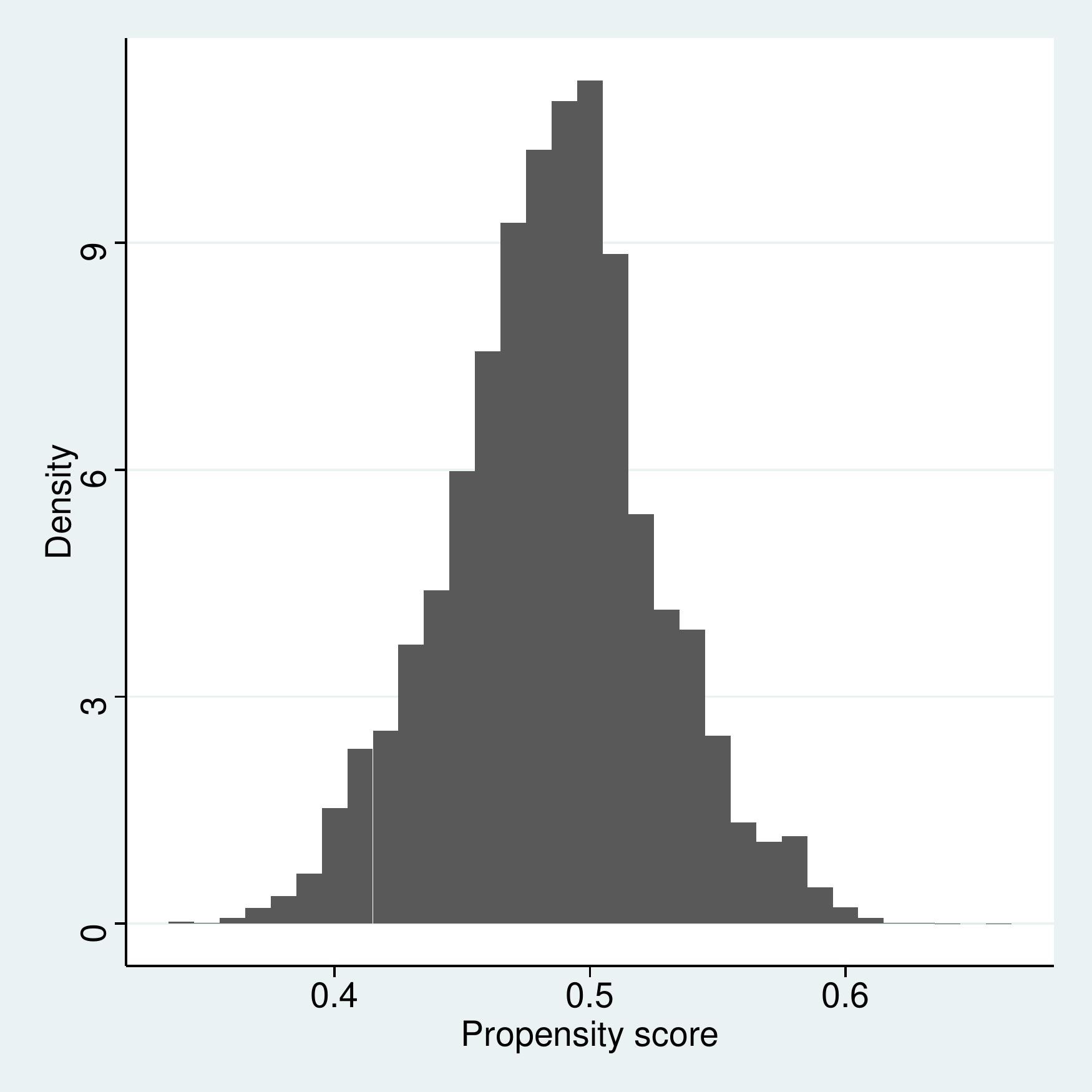}
    \caption{$\pm$ 1 p.p.}
\end{subfigure}
    \caption{Distribution of estimated propensity scores} 
    \label{fig:propensity}
    \caption*{\footnotesize{Note: Histogram plot of estimated treatment propensity scores: (a) for our benchmark analysis with vote margin interval of $\left[-5.0\%, +5.0\%\right]$; (b) for vote margin interval of $\left[-2.5\%, +2.5\%\right]$; and (c) for vote margin interval of $\left[-1.0\%, +1.0\%\right]$}}
\end{figure}

\begin{figure}[h!]
\small
\begin{subfigure}{.5\textwidth}
    \centering
        \includegraphics[scale=0.3]{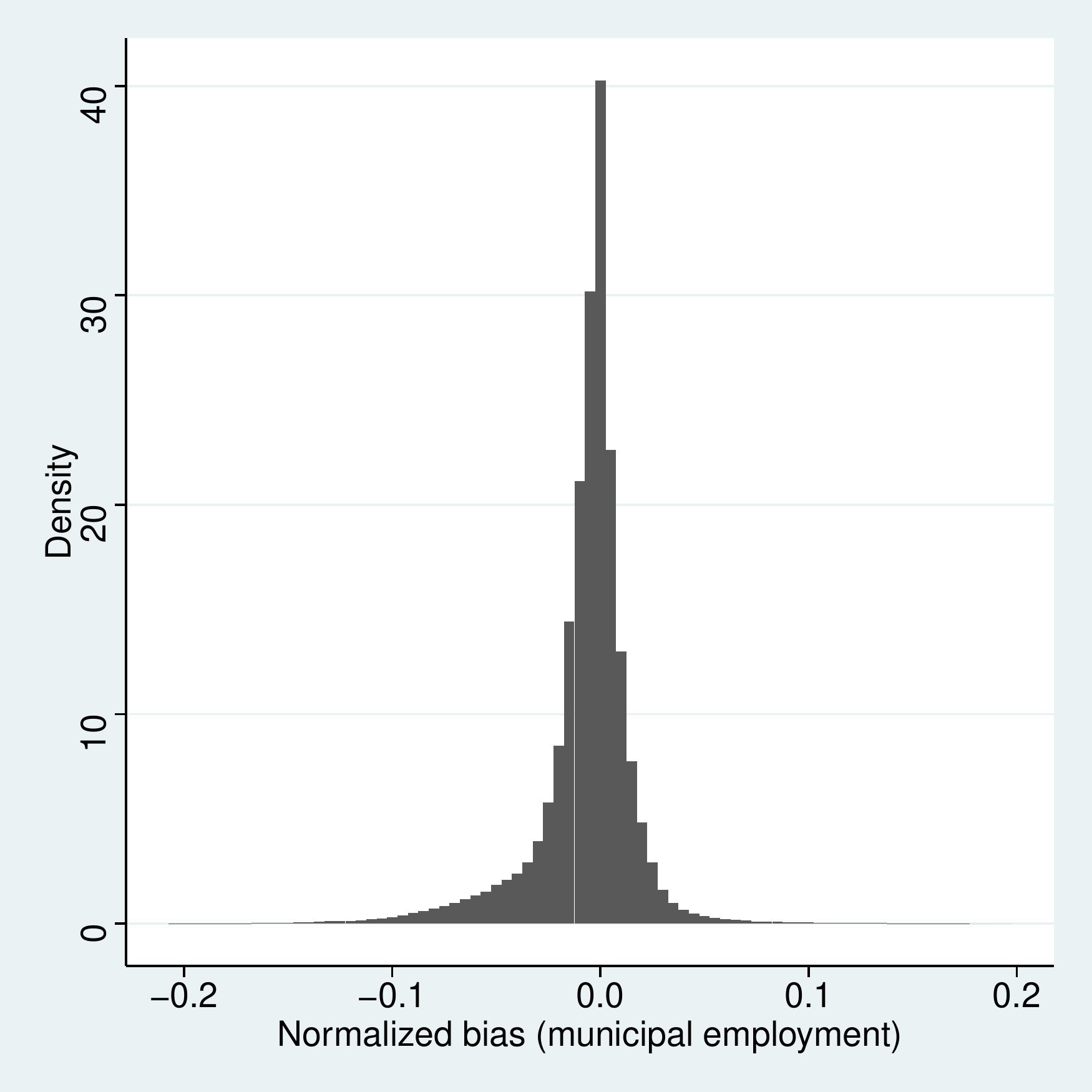}
    \caption{Municipal employment}
\end{subfigure}
\begin{subfigure}{.5\textwidth}
    \centering
        \includegraphics[scale=0.3]{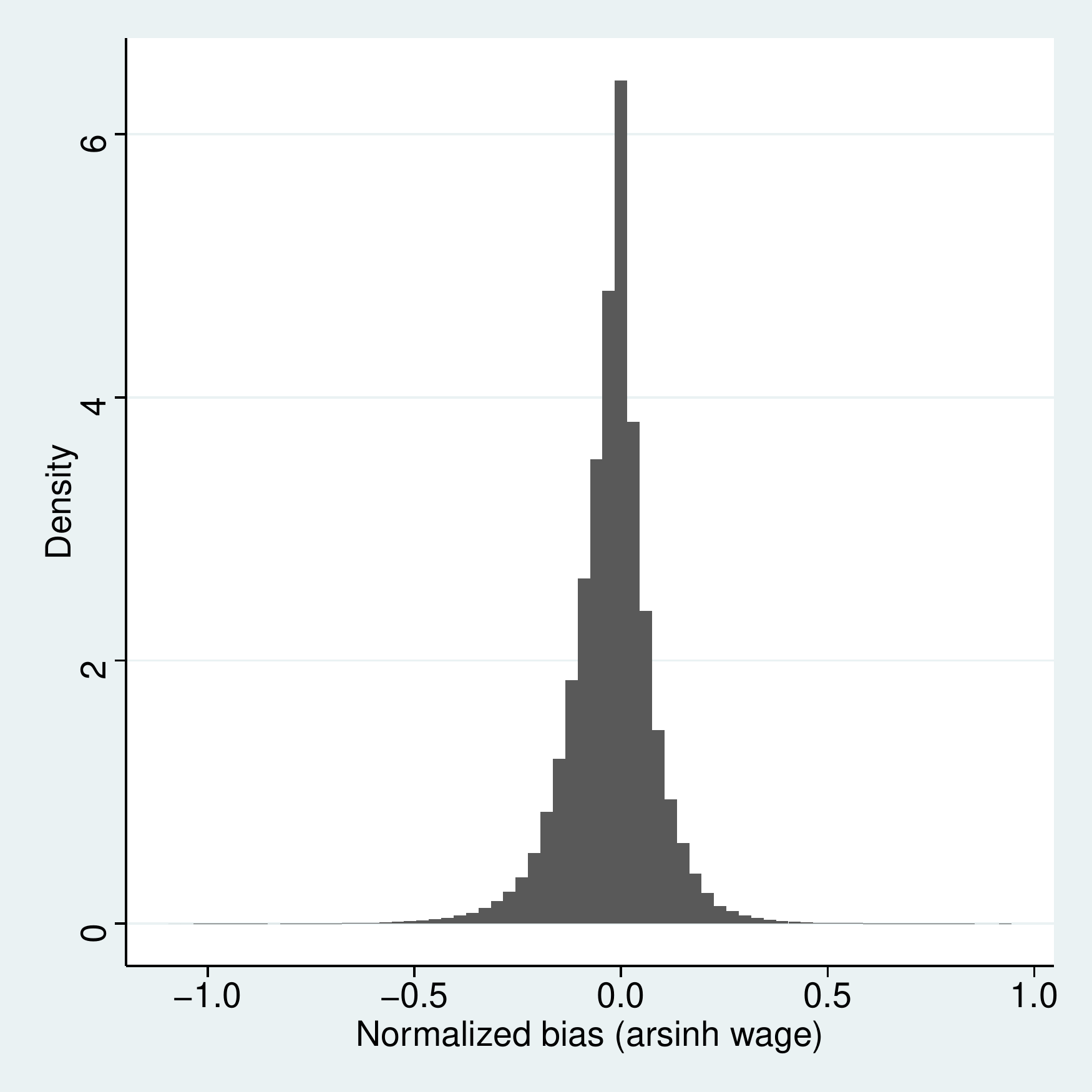}
    \caption{Employment wage (arsinh)}
\end{subfigure}
\begin{subfigure}{.5\textwidth}
    \centering
        \includegraphics[scale=0.3]{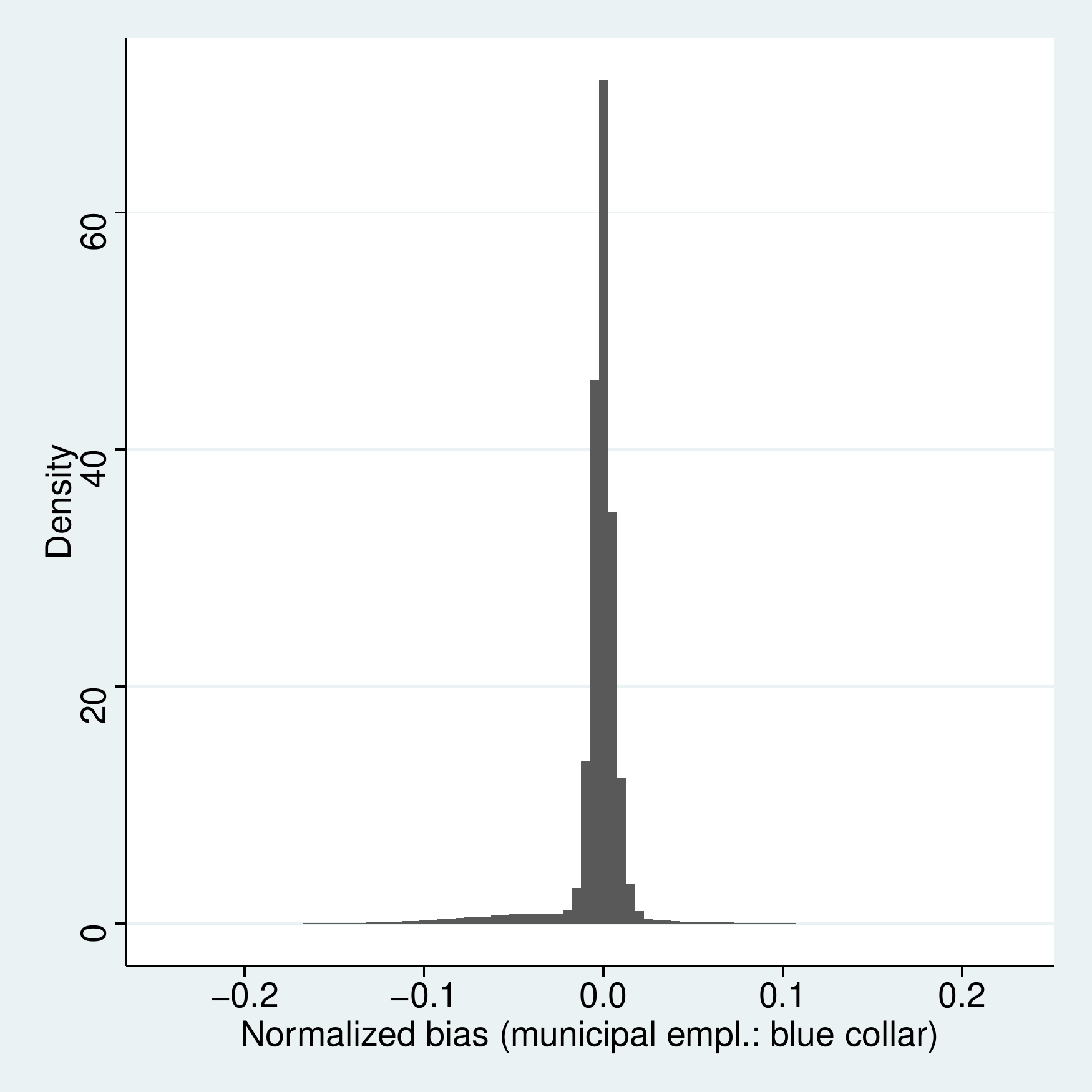}
    \caption{Municipal employment: blue collar}
\end{subfigure}
\begin{subfigure}{.5\textwidth}
    \centering
        \includegraphics[scale=0.3]{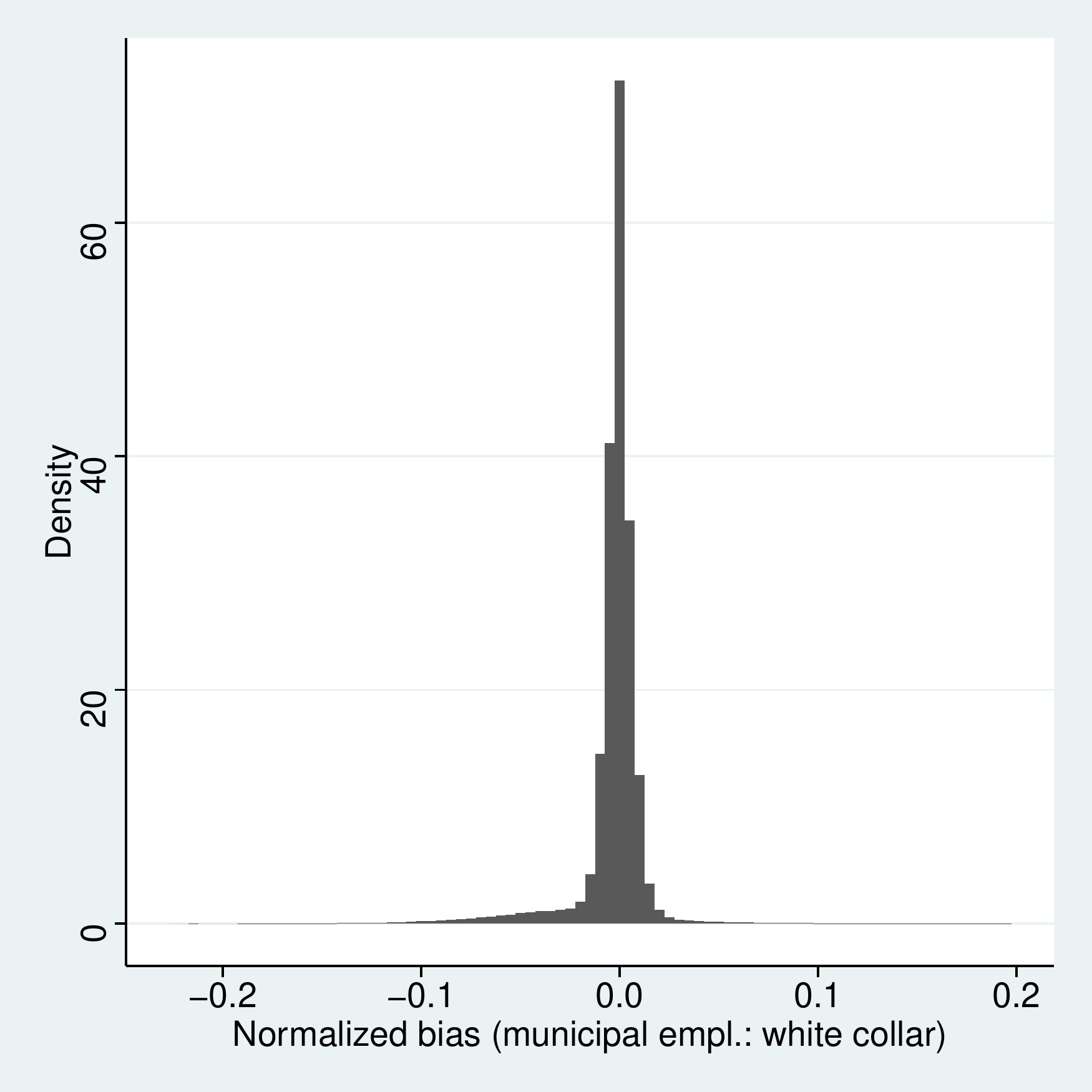}
    \caption{Municipal employment: white collar}
\end{subfigure}
\begin{subfigure}{\textwidth}
    \centering
        \includegraphics[scale=0.3]{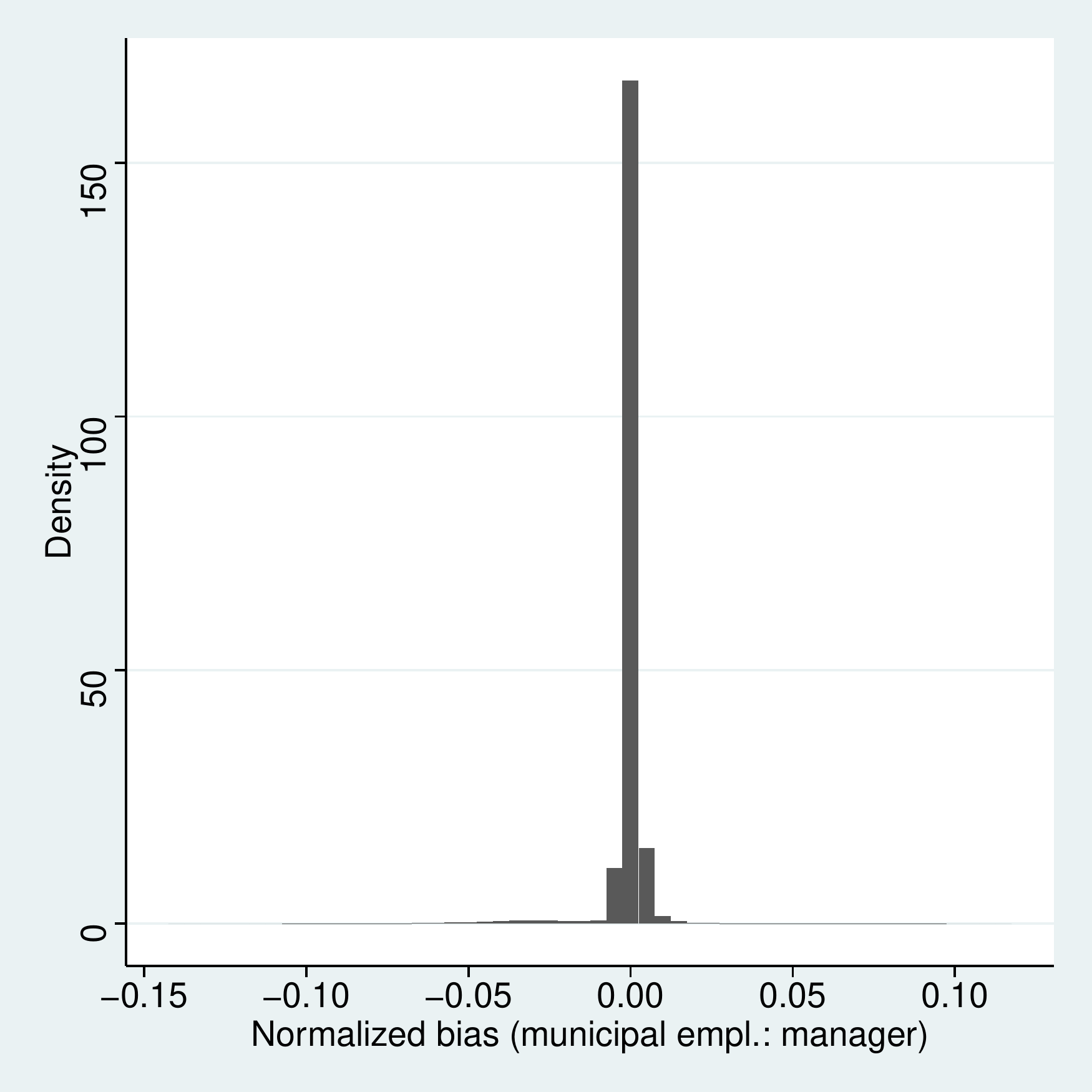}
    \caption{Municipal employment: manager}
\end{subfigure}
\caption{Estimated prediction bias for different dependent variables}
\caption*{\footnotesize{Note: Histogram plot of estimated biases for each outcome variable, according to equation $\text{bias}(x) = \left( W(x) - \mathbb{E}[W_i] \right) \left( \mathbb{E}[W_i] (\mu(0,x) - \mu(0)) + (1-\mathbb{E}[W_i]) (\mu(1, x) - \mu(1)) \right)$, divided by each outcome's standard deviation. (See the main text for further discussion.)}}
\label{fig:bias_full}
\end{figure}

I also follow \cite{athey2017estimating} and estimate the bias corrected by the propensity score as: $$b(x) = \left( W(x) - \mathbb{E}[W_i] \right) \left( \mathbb{E}[W_i] (\mu(0,x) - \mu(0)) + (1-\mathbb{E}[W_i]) (\mu(1, x) - \mu(1)) \right),$$ where $\mu(w, x) \equiv \mathbb{E} \left[ Y_i(w) \left| \right. X_i = x \right]$ and $\mu(0) \equiv \mathbb{E} \left[ Y_i(0) \right]$. Again encouragingly, Figure \ref{fig:bias_full} shows that the estimated biases for our benchmark analysis, scaled by the standard deviation of each outcome, have a large mass at zero. Given these results, our benchmark analysis is done with a $\pm5$\% bandwidth in margin of victory, but results for $\pm2.5$\% and $\pm1$\% bandwidths, available in Appendix Tables \ref{table:main_025} and \ref{table:main_001}, are quantitatively very similar.

% --------------------
\section{Results} \label{sec:results}
% --------------------

\begin{figure}[h!]
\small
\begin{subfigure}{.5\textwidth}
    \centering
        \includegraphics[scale=0.3]{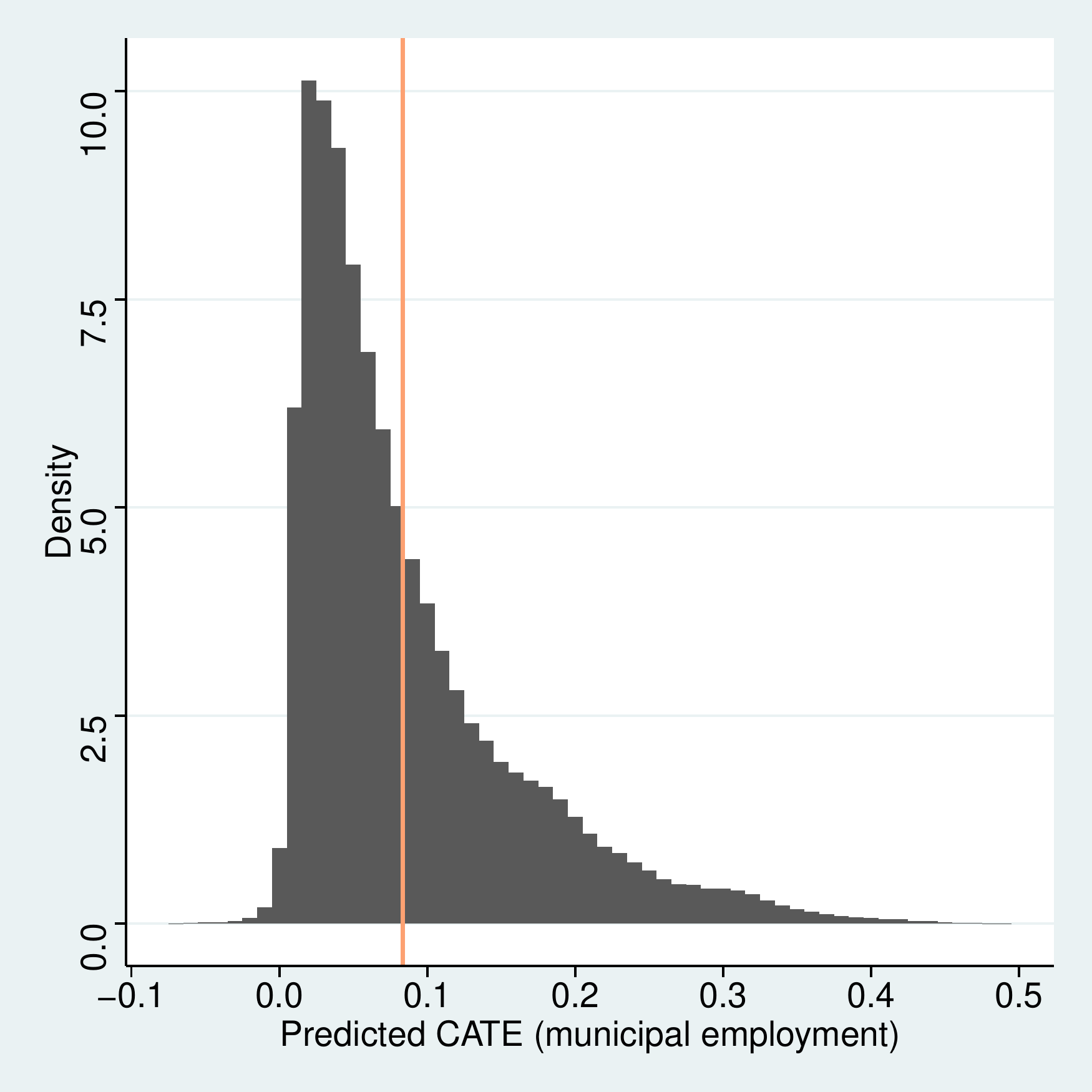}
    \caption{Municipal employment}
\end{subfigure}
\begin{subfigure}{.5\textwidth}
    \centering
        \includegraphics[scale=0.3]{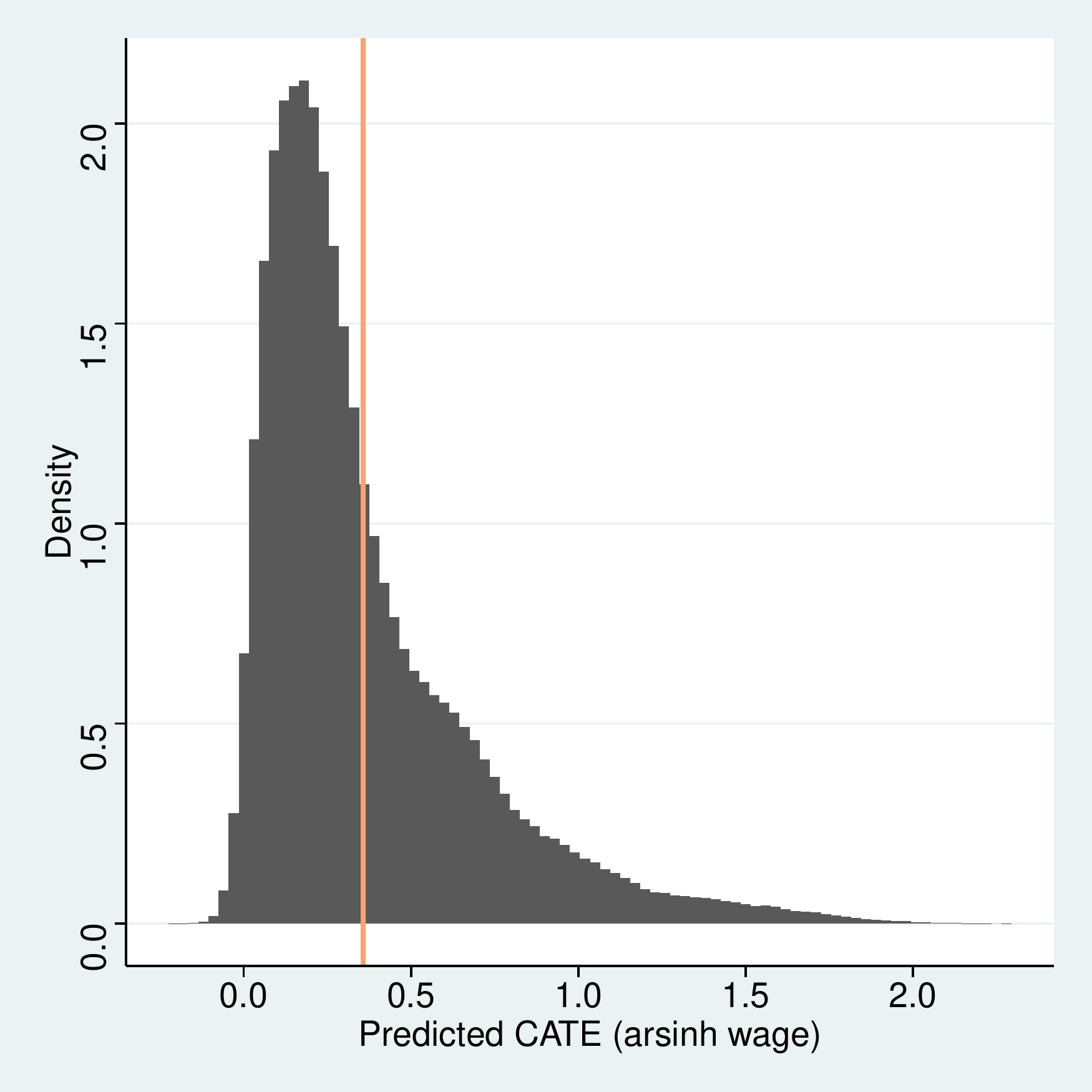}
    \caption{Employment wage (arsinh)}
\end{subfigure}
\begin{subfigure}{.5\textwidth}
    \centering
        \includegraphics[scale=0.3]{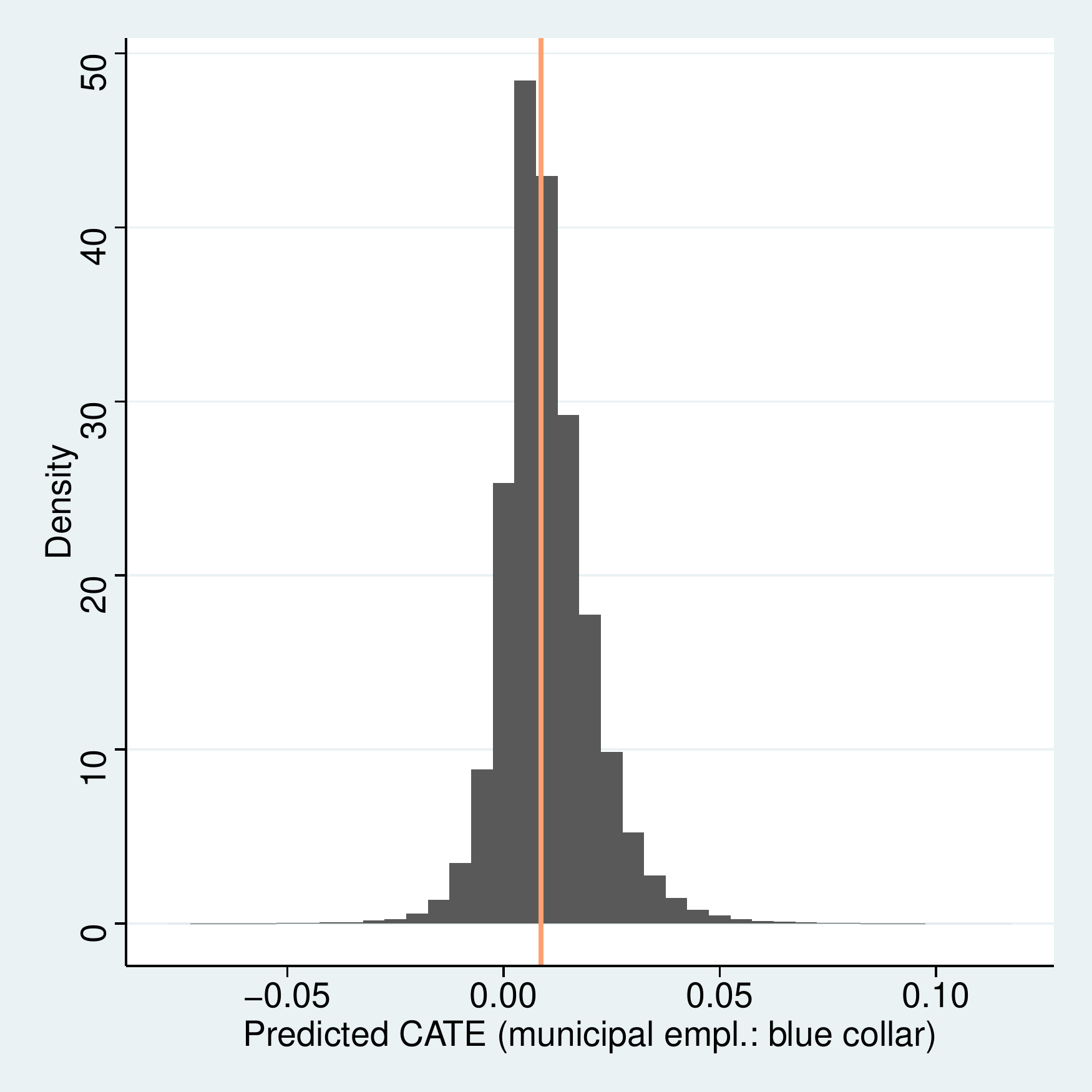}
    \caption{Municipal employment: blue collar}
\end{subfigure}
\begin{subfigure}{.5\textwidth}
    \centering
        \includegraphics[scale=0.3]{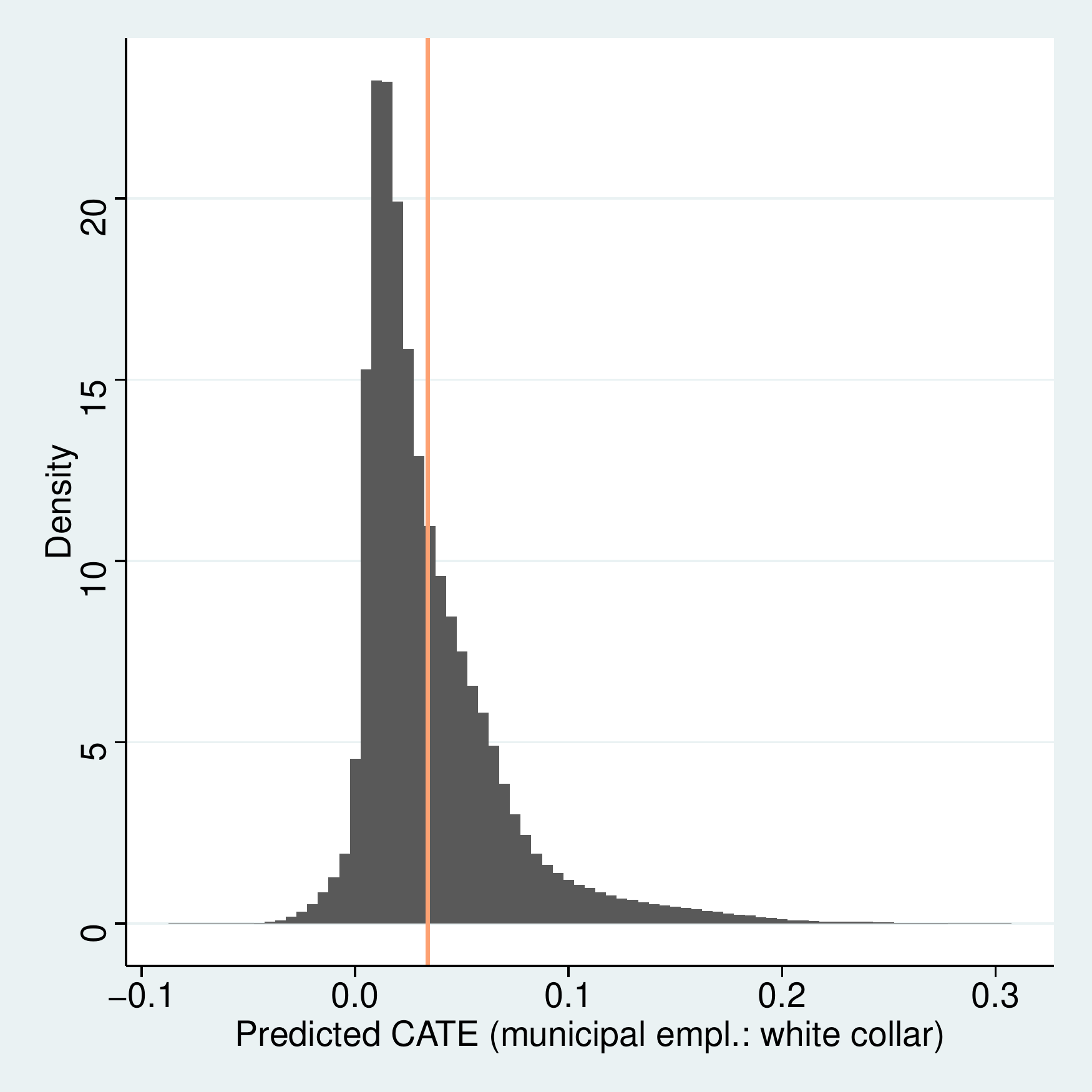}
    \caption{Municipal employment: white collar}
\end{subfigure}
\begin{subfigure}{\textwidth}
    \centering
        \includegraphics[scale=0.3]{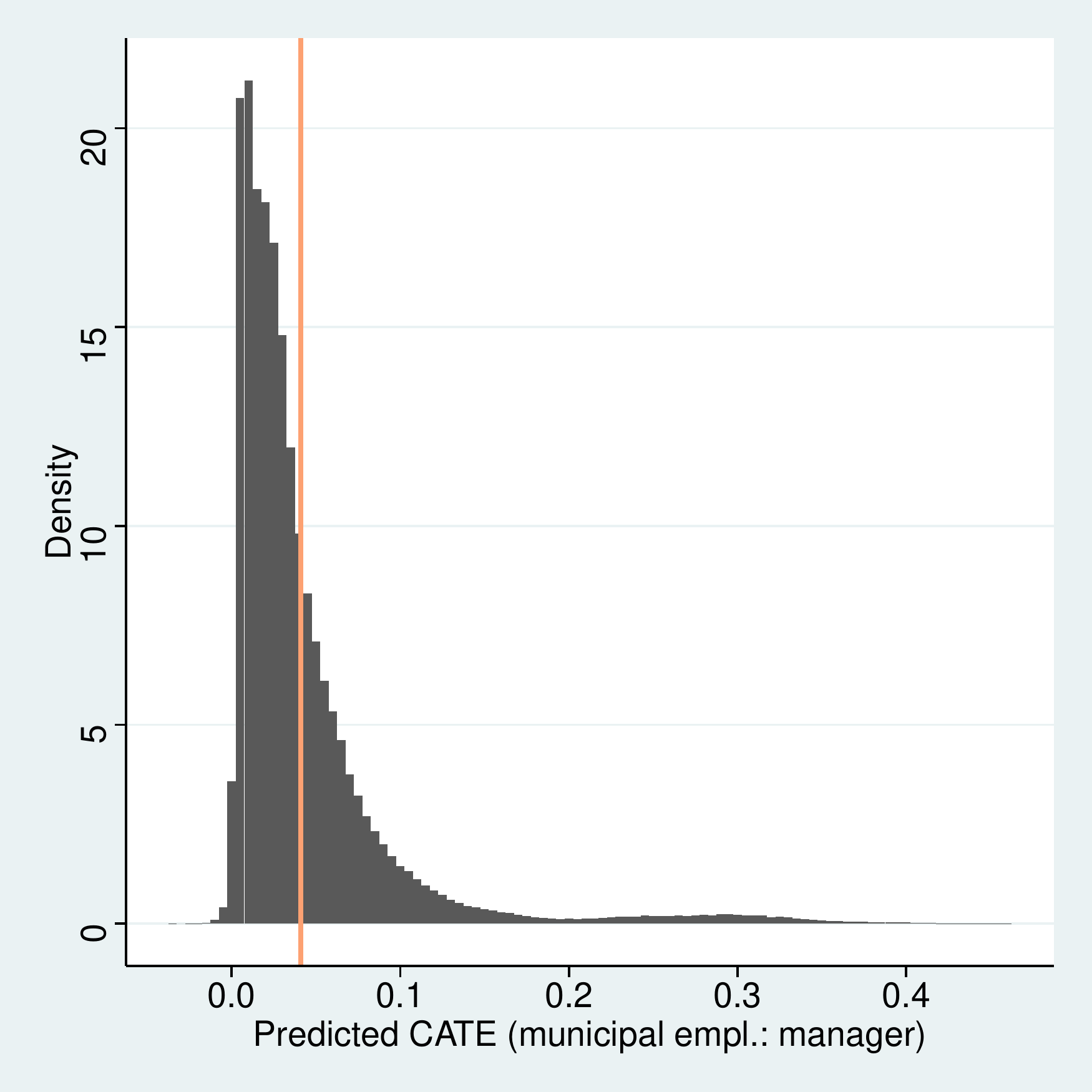}
    \caption{Municipal employment: manager}
\end{subfigure}
\caption{Predicted conditional average treatment effect for different dependent variables}
\caption*{\footnotesize{Note: Histogram plot of estimated conditional treatment effects for each of five outcomes: (a) being employed in the municipal public sector; (b) inverse hyperbolic sine of wages; (c) being employed in a blue collar municipal occupation; (d) being employed in a white collar municipal occupation; and (e) being employed in a managerial municipal occupation. The orange vertical line marks the average treatment effect for that outcome.}}
\label{fig:histogram_full}
\end{figure}

In Figure \ref{fig:histogram_full} we present the distribution of estimated conditional treatment effects for each outcome variable of interest. The orange line denotes the average treatment effect for that particular outcome. Although nowhere in the estimation procedure I restrict the conditional treatment effects to be positive, the estimated distribution is strongly censured at zero for almost all variables.\footnote{The only exception being blue collar employment, for which as we explain below the treatment effects are smaller and the prediction quality is poorer.} Since negative effects of political connections on municipal employment outcomes do not have a plausible theoretical (causal) explanation, this censoring is encouraging, as it suggests that we identify with precision the shape of the heterogeneity in treatment effects.

For most outcomes, the conditional treatment effect histogram resemble a exponential distribution, with large share of near zero coefficients, representing socioeconomic groups, parties and municipalities with small or no effect of political connections, but a long tail of large conditional treatment effects.

The main results are presented in Table \ref{table:main_full}, where I estimate best linear predictors of conditional treatment effects on all pre-treatment covariates, including term and municipality fixed effects and party dummies (these coefficients are omitted). Each column represents a generalized random forest model on a different outcome. 

Some notes on the interpretation of the table. Firstly, since all features are normalized, as explained in Section \ref{sec:data}, for continuous variables the interpretation of the coefficient is the best linear prediction of the effect of one standard deviation increase in the covariate on the conditional treatment effect. (The standard deviation of all covariates are available in Table \ref{tab:descriptive}.) Secondly, for the same reason, the coefficient of the regression intercept identifies the average treatment effect. Thirdly, I re-scale dummy variable coefficients (and standard errors) by the covariate standard deviation to return to a 0-1 interpretation usual to economics. However, it is important to note that since the intercept identifies the average treatment effect, unlike usual linear regression results, it does not represent the conditional treatment effect for the baseline (dummy equals zero). Therefore, to understand the linear projection of conditional treatment effects on a dummy group, we need to proportion the coefficients by the supplement of that group ($1 - p$ if $p$ is the share of individuals on that group).

\begin{table}[h!] \centering 
\begin{threeparttable}
\scriptsize
\caption{Best linear predictor of conditional treatment effect estimates} 
\label{table:main_full} 
\begin{tabular}{@{\extracolsep{5pt}}lccccc} 
\\ \hline 
\hline \\[-1.8ex] 
 & \multicolumn{5}{c}{\textit{Dependent variable:}} \\ 
\cline{2-6} 
\\[-1.8ex] & (1) & (2) & (3) & (4) & (5)\\ 
\\[-1.8ex] & Municipal & Wage & Mun. empl.:  & Mun. empl.:  & Mun. empl.: \\ 
\\[-1.8ex] & employment &  (arsinh) & blue collar & white collar & manager \\ 
\hline \\[-1.8ex] 
Constant (ATE) & 0.083$^{}$ & 0.357$^{}$ & 0.009$^{}$ & 0.034$^{}$ & 0.041$^{}$ \\ 
& (0.002) & (0.008) & (0.001) & (0.001) & (0.001) \\ 
  & & & & & \\ 
Age & 0.025$^{}$ & 0.126$^{}$ & 0.002$^{}$ & 0.007$^{}$ & 0.016$^{}$ \\ 
  & (0.001) & (0.007) & (0.001) & (0.001) & (0.001) \\ 
  & & & & & \\ 
High-school & $-$0.069$^{}$ & $-$0.313$^{}$ & 0.009$^{}$ & $-$0.029$^{}$ & $-$0.049$^{}$ \\ 
incomplete  & (0.004) & (0.023) & (0.003) & (0.003) & (0.002) \\ 
  & & & & & \\  
University & 0.067$^{}$ & 0.471$^{}$ & $-$0.012$^{}$ & 0.009 & 0.074$^{}$ \\ 
  & (0.008) & (0.043) & (0.004) & (0.007) & (0.005) \\ 
  & & & & & \\ 
Male & 0.050$^{}$ & 0.234$^{}$ & 0.005 & 0.016$^{}$ & 0.029$^{}$ \\ 
  & (0.005) & (0.026) & (0.003) & (0.004) & (0.003) \\ 
  & & & & & \\ 
Ability & $-$0.003$^{}$ & 0.005 & $-$0.004$^{}$ & $-$0.001 & 0.002$^{}$ \\ 
  & (0.001) & (0.007) & (0.001) & (0.001) & (0.001) \\ 
  & & & & & \\
Years since & $-$0.036$^{}$ & $-$0.149$^{}$ & $-$0.004$^{}$ & $-$0.016$^{}$ & $-$0.017$^{}$ \\ affiliation   & (0.002) & (0.008) & (0.001) & (0.001) & (0.001) \\ 
  & & & & & \\ 
Newly affiliated & 0.047$^{}$ & 0.049 & 0.033$^{}$ & 0.006 & 0.002 \\ 
  & (0.025) & (0.134) & (0.012) & (0.017) & (0.014) \\ 
  & & & & & \\  
\multicolumn{6}{@{\extracolsep{5pt}}l}{\textit{Previous employment characteristics:}} \\ 
   & & & & & \\ 
Blue collar & $-$0.074$^{}$ & $-$0.288$^{}$ & 0.020$^{}$ & $-$0.081$^{}$ & $-$0.007 \\ 
  & (0.008) & (0.043) & (0.009) & (0.009) & (0.004) \\ 
  & & & & & \\ 
Manager & 0.503$^{}$ & 2.409$^{}$ & $-$0.014 & $-$0.224$^{}$ & 0.761$^{}$ \\ 
  & (0.038) & (0.182) & (0.015) & (0.028) & (0.043) \\ 
  & & & & & \\
Government & 0.126$^{}$ & 0.854$^{}$ & 0.010 & 0.062$^{}$ & 0.054$^{}$ \\ 
  & (0.012) & (0.076) & (0.009) & (0.008) & (0.006) \\ 
  & & & & & \\ 
Employed & $-$0.041$^{}$ & $-$0.348$^{}$ & $-$0.014$^{}$ & 0.031$^{}$ & $-$0.063$^{}$ \\ 
  & (0.010) & (0.056) & (0.007) & (0.008) & (0.006) \\ 
  & & & & & \\
Tenure & $-$0.040$^{}$ & $-$0.163$^{}$ & $-$0.005$^{}$ & $-$0.022$^{}$ & $-$0.013$^{}$ \\ 
  & (0.002) & (0.009) & (0.001) & (0.001) & (0.001) \\ 
  & & & & & \\
Establishment & 0.012$^{}$ & 0.078$^{}$ & 0.001 & 0.006$^{}$ & 0.004$^{}$ \\ 
size  & (0.003) & (0.019) & (0.001) & (0.002) & (0.002) \\ 
  & & & & & \\ 
\hline \\[-1.8ex] 
Mean forest & 0.958 & 0.972 & 0.868 & 0.964 & 0.956 \\ 
prediction  & (0.021) & (0.023) & (0.093) & (0.036) & (0.024) \\ 
Differential forest & 1.234 & 1.248 & 1.538 & 1.411 & 1.147 \\ 
prediction  & (0.023) & (0.023) & (0.125) & (0.047) & (0.040) \\ 
N & 620,864 & 620,864 & 620,864 & 620,864 & 620,864 \\ 
  & & & & & \\ 
\hline 
\hline \\[-1.8ex] 
\end{tabular} 
\begin{tablenotes}
\item \emph{Notes:} All regressions include municipality and term fixed effects, as well as party dummies. Considers only municipal elections with vote margin within $\left[-5.0\%, +5.0\%\right]$. The dependent variables capture the employment situation in December of the first year at office. This regression finds the best linear predictor to the conditional average treatment effect estimates of the random forest model, weighting the OLS with the estimated propensity scores. Since covariates are demeaned, the regression constants identify the average treatment effects. Coefficients (and standard errors) of dummy variables are then adjusted for a binary variable interpretation. Errors are clustered at the municipality level.
\end{tablenotes}
\end{threeparttable}
\end{table} 

As aforementioned, the coefficient of the regression constant identifies the average treatment effect, and in the first row of Table \ref{table:main_full} we see that being politically connected to the mayor's party causally increases municipal employment by 8.3 percentages points, a 39.5\% increase over the baseline (21\% of our data are municipal employees). This increase is very heterogeneous across occupations, however. While for managerial positions the effect is a 256\% increase over the 1.6\% baseline, for white collar occupations it is a 35.4\% increase and for blue collar occupations just a 9.3\% increase over a baseline of 9.7\%. Indeed, while municipal workers in managerial positions account for only 14.6\% of municipal employees in our sample, they respond for 49\% of the effect of political connections on municipal employment (4.1 out of the 8.3 p.p. effect).

We also find that being a mayor's supporter increases wages earned by 42.9\%. Importantly, this includes the intensive margin effect of higher wages for already employed workers (in public and private sector), as well as the extensive margin effect of higher municipal employment, either by excess hirings or by affiliates to the losing candidate leaving public employment.\footnote{Here I employ the usual approximation of the inverse hyperbolic sine by the natural logarithm, see \cite{bellemare2020elasticities} for a discussion.} To better disentangle these margins, we can separately calculate this increase for each subgroup based on previous employment. Indeed, we find that this 42.9\% increase is the average effect of a striking increase of 81.6\% in earnings of affiliates in the public sector, a 47.8\% increase in wages of previously unemployed workers (compared to control group), and a modest 6.2\% rise in wages for workers previously in the private sector.\footnote{The last result is available in Table \ref{table:private}, while the first two are available in Appendix Tables \ref{table:main_public} and \ref{table:main_nowork}, respectively.}

As measures of quality of fit, in Table \ref{table:main_full} we present the mean forest prediction and differential forest prediction for each of our causal forests. A mean forest prediction coefficient of one suggests the mean forest prediction is correct, and a differential forest prediction coefficient of one means the causal forest captures well the heterogeneity on the conditional treatment effects \citep{athey2017estimating}. We observe that for most of our causal forests the mean forest prediction is very close to one, in some cases statistically indistinguishable from it. The exception is the municipal blue collar employment regression, which has ATE closer to zero and is less well identified. All causal forests have high differential forest prediction.

\paragraph{Best linear predictor of conditional treatment effects.}

Our main contribution, however, is to investigate heterogeneity in the effect of being affiliated to the mayor's party on municipal employment conditional on worker characteristics. In the second row, we see that the conditional treatment effect on municipal employment is 2.5 p.p. higher when comparing affiliates to the mayor's party and second-place candidate that are one standard deviation older (11.3 years) from the mean. Political connections are also more valuable for men: while women have a 5 percentage points higher chance of being employed in the municipal public sector if they are affiliated to the mayor's party, as opposed to the losing candidate's, men have a 10 p.p. higher probability.\footnote{Since 8.3 p.p. is the average treatment effect, and 65\% of our sample is male, which have a 5 p.p. higher conditional treatment effect (Table \ref{table:main_full}, 5th row), we reach these numbers by a rule of three calculation.}

An important heterogeneity to consider is whether political connections are more relevant for low or high education workers. If low education workers were substantially less likely to be employed in municipal government, and especially at white collar and managerial positions, except when politically connected, this would represent convincing evidence that patronage acts towards reducing bureaucratic efficiency, as it increases the share of low education workers in the municipal government.

What we observe, however, is the opposite. While about one-forth of municipal employees in managerial positions in our data have incomplete high-school (or less) education, the conditional treatment effect of being affiliated to the mayor's party in this sub-sample is essentially zero (-0.001), while for the university educated it is estimated as 12.1 percentage points (a striking 715\% increase over the baseline). 

Looking at descriptive statistics, we see that for affiliates of the losing candidates, 2.8\% of those with higher education work at municipal managerial positions, while 1.4\% of those without higher education do. On the other hand, among those affiliated to the winning candidates, 4.3\% without higher education work at municipal management positions, and 10.3\% of affiliates with university education do. Clearly, high education workers are more likely to work in the municipality management, but this difference is \textit{larger} for politically connected workers, the opposite of what one would expect if patronage were negatively selecting on education. Our heterogeneity analysis of CATE is a more formal and careful way of making the same argument.

Here I note that a common objection to this analysis is that municipalities which hire low education workers for managerial public positions are different than municipalities that hire high education workers (the latter being probably bigger, richer, and better geographically positioned), which could be correlated to the importance of political connections in the municipality. However, unlike previous research, our method is able to control for municipality fixed-effects both in our treatment effect estimation procedure, as well as in our heterogeneity analysis, dealing with such concerns. In this sense, while still not quasi-experimental, our heterogeneity analysis is more convincing than previous research.

The same result is found for white collar occupations, where the treatment effect for incomplete high-school workers (15.8\% of white-collar municipal employees) is 1.54 percentage points, smaller than for workers with complete high-school at 4.4 p.p., and especially with university education at 5.34 p.p., a 314.6\% increase over the baseline for this subgroup. The small effect of political connections on low education workers might be due to improving institutional features and an overall increase in educational attainment in Brazil, which led to increasing difficulty in hiring low education employees for white collar occupations (especially school teachers and health workers). Indeed, we observe that the share of municipal white collar employees without at least complete high-school fell by 37\% in the 12 years covered by our analysis.

As already mentioned, this result contradicts previous research, that pointed towards negative selection on education. This difference results from my analysis being able to investigate heterogeneity regarding all features jointly. For example, since university educated affiliates are more likely to be previously employed (in our data, 53.2\% versus 45.8\% for those without university degree), and the conditional treatment effect is larger for previously unemployed workers, as I note below, this would bias downward the positive heterogeneity effect of high education on the value of political connection in their analysis (but not mine).

Only for blue collar occupations we observe the opposite effect; namely, that the conditional treatment effect of political connections is larger for workers with incomplete high-school and smaller for highly educated workers. This is intuitive, since blue collar municipal employment usually involves hard work and low pay, being unappealing for the highly educated. In fact, we observe that in our sample only 6.4\% of these employees have university education. 

Accordingly, as opposed to previous literature, I find that political connections have very little importance for blue collar occupations. Indeed, when looking at descriptive statistics, we see that, in our data (that is, in close elections), workers affiliated to the losing candidate are (very slightly) more likely to work in public blue collar occupations than workers affiliated to the mayor's party (9.7\% versus 9.52\%). When controlling for previous characteristics, our causal forest estimates a positive effect, but economically insignificant. I conjecture that these jobs are not appealing for patronage appointments, as they pay too little, but this result is also consistent with political connections being used to select workers particularly for policy meaningful positions.

The next aspect on which we study the heterogeneous effects of political connections is previous employment characteristics. Our first finding is that unemployed (or informally employed) workers and public sector employees are the ones that benefit the most from political connections. Since informal work usually has low pay and no benefits, unemployed or informally employed workers have the most to gain from public employment. This has two effects. On the one hand, since the economic rent to be bargained is larger, buying the support of these workers with patronage is cheaper for political candidates, and they are more likely to do it. On the other hand, these workers are more likely to be employed by the mayor also for ideological reasons, since the unemployed are more willing to accept municipal work (even if it is not patronage), because they have worse outside options.

Moreover, we find that mayors hire affiliates from their own party significantly more often when these affiliates were previously employed in the same occupations they are hired to in municipal government. Even for municipal managerial positions, where only half of previously employed workers were also employed in management, we see a large positive heterogeneous effect.\footnote{One disadvantage of non-standardized dummy coefficients is that when they represent a small share of the sample (4.5\% in the case of workers previously in management), the coefficient is inflated and loses some interpretability. The sign of the coefficient remains valid, however.} We discuss these results further when analyzing Table \ref{table:private} below.

Although important, education is not the only measure of worker quality, and we would like to understand whether political connections negatively select on ability. As explained in Section \ref{sec:data}, we construct a proxy for ability by estimating Mincerian individual fixed-effects using our long panel of formal sector earnings. We then use this variable in our causal forest estimation and heterogeneity analysis. Unlike education, we observe that political connection is less beneficial for high ability workers. 

Interestingly, in a sample of affiliates previously employed in private sector positions (Table \ref{table:private}), we see a larger effect: one standard deviation higher ability corresponds to a 0.9 percentage points lower chance of municipal employment, over 34\% of the ATE. The interpretation of these results, however, is ambiguous. On the one hand, public sector appointments are more valuable for low ability workers, and therefore it is a more efficient method of patronage. On the other hand, even good quality workers would be less willing to join the public bureaucracy if their outside options on the private sphere are more attractive.

Finally, another relevant contribution of this research is to investigate whether recently affiliated workers benefit more from being politically connected to the mayor than affiliates with longer tenure. This is important, because affiliates with long party tenure are more strongly connected to the party bureaucracy, while new party members are plausibly affiliated for pragmatic reasons.\footnote{One could be concerned that since political parties are different in age, this could bias our heterogeneity results. While this concern is valid in traditional heterogeneity analysis, our best linear predictor of CATE measures heterogeneity in treatment effect within parties.} And indeed we find that a standard deviation increase in party tenure (8.5 years) is associated with 3.6 percentage points smaller effect of political patronage, or 43.3\% of ATE.

Moreover, we test whether there is a differential effect of being affiliated the year before the election. The year before election in Brazil is when parties choose the mayoral candidates and distribute party financing. Newly affiliates, therefore, might join the party to influence party democracy and help with the election campaign, and if political connections are used as patronage, these workers would be more likely to be rewarded with public jobs after the election. Although the estimate is very noisy, since it relies on the linearity of the years since affiliation coefficient, this is what we observe, with newly affiliated workers having 4.4 percentage points higher treatment effect of being affiliated to the mayor's party on the probability of municipal employment.

\paragraph{Classification analysis of heterogeneous treatment effects.}

To better understand the heterogeneous treatment effects of political connections, we employ classification analysis (CLAN; see \cite{chernozhukov2018generic}). We divide our sample into quartiles by estimated conditional treatment effect, and compare the characteristics of the most affected group with characteristics of the least affected group. Once again, we find that our features capture substantial heterogeneity: while the most affected quartile has a conditional treatment effect of being affiliated to the mayor's party of 0.21 (std. 0.004), the least affected group has economically insignificant effect of 0.01 (std. 0.002). 

We can also compare the mean features of each of these quartiles, which we do in Table \ref{tab:clan} below.\footnote{While I present standard deviations of all variables for completeness, it is noteworthy to remember that the standard deviation of binary variables, as trivially defined by $\sqrt{p(1-p)}$, is wholly uninformative.} We see that the most affected group is slightly younger, but much more educated: the share of the most affected group with university education is 4 times higher than in the least affected group (and double the sample average), while the share without complete high-school is half as big as for the lowest quartile.

They are very heterogeneous regarding previous employment as well. While two thirds of the most affected group are unemployed, and almost the entirety of the rest is working in government, with a sizable proportion is in managerial positions, the least affected group is almost entirely working, mostly on private sector and in blue collar occupations. They work for substantially longer tenure and in slightly larger companies.

Interestingly, we note that while the most affected group is more educated, they have lower than average ability (ability is a standardized variable), while the least affected group has almost half a standard deviation higher Mincerian fixed-effects on average.\footnote{Note that the Mincerian fixed-effects, our proxy for worker ability, are estimated orthogonal to education (and other observables).} Finally, the highest quartile on CATE is on average affiliated to the mayor's party for 4 years, with 11\% of them affiliated on the year before the election, while the lowest quartile is affiliated on average for 11 years, and only 1.8\% of them new affiliates.

% Table created by stargazer v.5.2.3 by Marek Hlavac, Social Policy Institute. E-mail: marek.hlavac at gmail.com
% Date and time: ter, abr 12, 2022 - 14:31:49
\begin{table}[!htbp] \centering 
\begin{threeparttable}
\small
\caption{Classification analysis of conditional treatment effect}
  \label{tab:clan} 
\begin{tabular}{@{\extracolsep{5pt}}lcc} 
\\[-1.8ex]\hline 
\hline \\[-1.8ex] 
 & \multicolumn{1}{c}{Most affected} & \multicolumn{1}{c}{Least affected} \\ 
 & group & group \\
\hline \\[-1.8ex] 
Age & 39.268 [11.655] & 44.062 [10.269] \\ 
High-school incomplete & 0.227 [0.419] & 0.564 [0.496] \\ 
University & 0.361 [0.480] & 0.094 [0.292] \\ 
Male & 0.627 [0.484] & 0.681 [0.466] \\ 
Blue collar (lag) & 0.029 [0.167] & 0.680 [0.467] \\ 
Manager (lag) & 0.111 [0.314] & 0.008 [0.087] \\ 
Government job (lag) & 0.312 [0.463] & 0.293 [0.455] \\ 
Employed (lag) & 0.342 [0.474] & 0.835 [0.371] \\ 
Ability & $-$0.122 [0.560] & 0.413 [0.848] \\ 
Tenure (lag) & 0.926 [1.643] & 8.200 [8.561] \\ 
Estab. size (lag) & 2.532 [3.721] & 4.564 [3.225] \\ 
Years affiliated & 4.074 [4.480] & 11.656 [6.216] \\ 
Newly affiliated & 0.115 [0.319] & 0.018 [0.134] \\ 
Propensity score ($\hat{W}$) & 0.497 [0.052] & 0.482 [0.034] \\ 
Predicted outcome ($\hat{Y}$) & 0.282 [0.164] & 0.280 [0.383] \\ 
Treatment effect ($\hat{\tau}$) & 0.196 [0.066] & 0.019 [0.009] \\
Bias ($\hat{b}$) & $-$0.004 [0.009] & $-$0.004 [0.014] \\ 
\hline \\[-1.8ex] 
\end{tabular} 
\begin{tablenotes}
\item \emph{Notes:} Means and standard deviations (in brackets) of each feature within the most affected group (highest quartile) and least affected group (lowest quartile) of estimated conditional average treatment effects. Considers only municipal elections with vote margin within $\left[-5.0\%, +5.0\%\right]$. The dependent variables is whether employed in the municipality in December of the first year at office.
\end{tablenotes}
\end{threeparttable}
\end{table} 

\paragraph{Conditional treatment effect on workers previously employed in the private sector.}

In order to better discern between workers previously in different sectors of activity (or unemployed), in Table \ref{table:private} I present the best linear predictor of conditional treatment effects restricting the sample to workers previously employed in the private sector (23.8\% of the data).

For workers previously employed in the private sector, the average treatment effects are smaller, but they represent sizable increases over a minuscule baseline. The share of workers affiliated to the loser's party that were previously employed in the private sector and move to municipal employment is only 1\%, so a 2.6 percentage point increase corresponds to 260\% increase over the baseline.

Overall, we observe the same patterns as in our benchmark analysis regarding education, years since affiliation, and previous employment, but in some cases with larger variance. Interestingly, for this sub-sample the estimated heterogeneity on ability is still negative and everywhere significant.\footnote{Presumably, the results are stronger for this sub-sample for the mechanical fact that we are able to more precisely estimate the fixed-effects of the Mincerian equation.} One standard deviation increase in ability reduces the ATE in 34.6\%.

Table \ref{table:private} also clarifies that even for workers hired from the private sector, they tend to be hired by the municipality in occupations similar to what they were previously employed. The treatment effect of being affiliated to the mayor's party on municipal white collar employment is significantly smaller for workers previously employed in either blue collar or managerial occupations. Similarly, the CATE for public managerial positions is 0.37 percentage points for workers previously in blue collar positions, whereas for workers in white collar or managerial positions the CATE is 2 percentage points, a 80.75\% increase over the baseline.  

% Table created by stargazer v.5.2.2 by Marek Hlavac, Harvard University. E-mail: hlavac at fas.harvard.edu
% Date and time: Fri, Oct 29, 2021 - 07:45:59 AM
\begin{table}[h!] \centering 
\begin{threeparttable}
\scriptsize
\caption{Best linear predictor of conditional treatment effect estimates -- previous job in private sector} 
\label{table:private} 
\begin{tabular}{@{\extracolsep{5pt}}lccccc} 
\\ \hline 
\hline \\[-1.8ex] 
 & \multicolumn{5}{c}{\textit{Dependent variable:}} \\ 
\cline{2-6} 
\\[-1.8ex] & (1) & (2) & (3) & (4) & (5)\\ 
\\[-1.8ex] & Municipal & Wage & Mun. empl.:  & Mun. empl.:  & Mun. empl.: \\ 
\\[-1.8ex] & employment &  (arsinh) & blue collar & white collar & manager \\ 
\hline \\[-1.8ex] 
Constant (ATE) & 0.026$^{}$ & 0.061$^{}$ & 0.004$^{}$ & 0.011$^{}$ & 0.011$^{}$ \\ 
  & (0.001) & (0.013) & (0.0005) & (0.001) & (0.001) \\ 
  & & & & & \\ 
Age & 0.009$^{}$ & 0.041$^{}$ & 0.002$^{}$ & 0.002$^{}$ & 0.005$^{}$ \\ 
  & (0.001) & (0.015) & (0.001) & (0.001) & (0.001) \\ 
  & & & & & \\ 
High-school& $-$0.027$^{}$ & $-$0.036 & $-$0.001 & $-$0.013$^{}$ & $-$0.012$^{}$ \\ 
incomplete  & (0.004) & (0.049) & (0.002) & (0.002) & (0.002) \\ 
  & & & & & \\ 
University & 0.089$^{}$ & 0.202$^{}$ & 0.001 & 0.042$^{}$ & 0.046$^{}$ \\ 
  & (0.012) & (0.101) & (0.004) & (0.008) & (0.007) \\ 
  & & & & & \\ 
Male & 0.001 & $-$0.011 & 0.001 & $-$0.005$^{}$ & 0.004 \\ 
  & (0.005) & (0.057) & (0.003) & (0.003) & (0.003) \\ 
  & & & & & \\ 
Ability & $-$0.009$^{}$ & $-$0.037$^{}$ & $-$0.002$^{}$ & $-$0.004$^{}$ & $-$0.004$^{}$ \\ 
  & (0.001) & (0.013) & (0.001) & (0.001) & (0.001) \\ 
  & & & & & \\ 
Years since & $-$0.016$^{}$ & $-$0.044$^{}$ & $-$0.003$^{}$ & $-$0.006$^{}$ & $-$0.007$^{}$ \\ 
affiliation & (0.001) & (0.015) & (0.001) & (0.001) & (0.001) \\ 
  & & & & & \\  
Newly affiliated & 0.028 & $-$0.119 & 0.012 & 0.006 & 0.006 \\ 
  & (0.022) & (0.233) & (0.011) & (0.014) & (0.011) \\ 
  & & & & & \\ 
\multicolumn{6}{@{\extracolsep{5pt}}l}{\textit{Previous employment characteristics:}} \\ 
   & & & & & \\ 
Blue collar & $-$0.020$^{}$ & $-$0.078 & 0.004 & $-$0.008$^{}$ & $-$0.017$^{}$ \\ 
  & (0.006) & (0.060) & (0.003) & (0.004) & (0.004) \\ 
  & & & & & \\ 
Manager & $-$0.034$^{}$ & $-$0.362 & $-$0.006 & $-$0.027$^{}$ & 0.001 \\ 
  & (0.020) & (0.250) & (0.006) & (0.013) & (0.014) \\ 
  & & & & & \\
Tenure & $-$0.005$^{}$ & $-$0.018 & $-$0.001$^{}$ & $-$0.001$^{}$ & $-$0.002$^{}$ \\ 
  & (0.001) & (0.012) & (0.0004) & (0.001) & (0.0004) \\ 
  & & & & & \\ 
Establishment & $-$0.008$^{}$ & $-$0.016 & $-$0.001 & $-$0.003$^{}$ & $-$0.004$^{}$ \\ 
size  & (0.001) & (0.014) & (0.001) & (0.001) & (0.001) \\ 
  & & & & & \\  
\hline \\[-1.8ex] 
%Mean debiased error & 0.030 & 2.872 & 0.009 & 0.013 & 0.008 \\ 
%MDE / treated mean & 65.0\% & 80.62\% & 88.1\% & 67.1\% & 55.3\% \\ 
Mean forest & 0.916 & 0.996 & 0.881 & 0.927 & 0.910 \\ 
prediction  & (0.043) & (0.205) & (0.117) & (0.056) & (0.054) \\ 
Differential forest & 1.287 & 0.395 & 0.955 & 1.301 & 1.257 \\ 
prediction  & (0.105) & (0.189) & (0.178) & (0.134) & (0.108) \\ 
N  & 148,308 & 148,308 & 148,308 & 148,308 & 148,308 \\ 
  & & & & & \\ 
\hline 
\hline \\[-1.8ex] 
\end{tabular} 
\begin{tablenotes}
\item \emph{Notes:} All regressions include municipality and term fixed effects, as well as party dummies. Considers only municipal elections with vote margin within $\left[-5\%, +5\%\right]$. Sample restricted to individuals with private formal jobs in the previous year. The dependent variables are the values in the first year at office. This regression finds the best linear predictor to the conditional average treatment effect estimates of the random forest model, weighting the OLS with the estimated propensity scores. Since covariates are demeaned, the regression constant identifies average treatment effects. Errors are clustered at the municipal level.
\end{tablenotes}
\end{threeparttable}
\end{table} 

% --------------------
\section{Conclusion}
% --------------------

This paper applies causal forests to better understand political influence in public sector appointments, by investigating heterogeneous effects of being affiliated to the mayor's party by worker characteristics. We find that more recent affiliates, and particularly those affiliated the year before elections, benefit the most from being affiliated to the mayor's party, indicating that these appointments are used as reward for political and campaign support (patronage), rather than simply indicating ideological proximity. But we do not find that these appointments are used to disproportionately place low education workers in the municipal bureaucracy. Rather, we find that political connections \textit{positively} select on education. However, since political connections select negatively on non-observable earning ability, the final impact of patronage on bureaucratic capacity is ambiguous.

Further research should seek to better understand the motivations behind political affiliation, especially given the knowledge that we uncover that newly affiliated supporters are much more likely to be given place in the municipal bureaucracy in case of their party's electoral victory.

% %%%%%%%%%%%%%%%%%%%%%%%%%%%%%%%%%%%%%%%%%%%%%%%%%%%%%%%%%%
% %%%%%%%%%%%%%%%%%%%%%%%%%%%%%%%%%%%%%%%%%%%%%%%%%%%%%%%%%%
% REFERENCES SECTION
% %%%%%%%%%%%%%%%%%%%%%%%%%%%%%%%%%%%%%%%%%%%%%%%%%%%%%%%%%%
% %%%%%%%%%%%%%%%%%%%%%%%%%%%%%%%%%%%%%%%%%%%%%%%%%%%%%%%%%%
\medskip

\bibliography{references.bib}

% %%%%%%%%%%%%%%%%%%%%%%%%%%%%%%%%%%%%%%%%%%%%%%%%%%%%%%%%%%
% %%%%%%%%%%%%%%%%%%%%%%%%%%%%%%%%%%%%%%%%%%%%%%%%%%%%%%%%%%
% TABLES
% %%%%%%%%%%%%%%%%%%%%%%%%%%%%%%%%%%%%%%%%%%%%%%%%%%%%%%%%%%
% %%%%%%%%%%%%%%%%%%%%%%%%%%%%%%%%%%%%%%%%%%%%%%%%%%%%%%%%%%

% Table created by stargazer v.5.2.2 by Marek Hlavac, Harvard University. E-mail: hlavac at fas.harvard.edu
% Date and time: Fri, Oct 29, 2021 - 07:45:59 AM

% %%%%%%%%%%%%%%%%%%%%%%%%%%%%%%%%%%%%%%%%%%%%%%%%%%%%%%%%%%
% %%%%%%%%%%%%%%%%%%%%%%%%%%%%%%%%%%%%%%%%%%%%%%%%%%%%%%%%%%
% FIGURES
% %%%%%%%%%%%%%%%%%%%%%%%%%%%%%%%%%%%%%%%%%%%%%%%%%%%%%%%%%%
% %%%%%%%%%%%%%%%%%%%%%%%%%%%%%%%%%%%%%%%%%%%%%%%%%%%%%%%%%%

%\begin{subfigure}{.5\textwidth}
%    \centering
%        \includegraphics[scale=0.4]{figures/propensity_score_full_incumbent.pdf}
%    \caption{With incumbents}
%\end{subfigure}

%\begin{figure}[H]
%    \centering
%        \includegraphics[scale=0.8]{figures/tree_y1_full.pdf}
%    \caption{Classification tree for conditional average treatment effects}
%    \label{fig:tree}
%\end{figure}

% %%%%%%%%%%%%%%%%%%%%%%%%%%%%%%%%%%%%%%%%%%%%%%%%%%%%%%%%%%
% %%%%%%%%%%%%%%%%%%%%%%%%%%%%%%%%%%%%%%%%%%%%%%%%%%%%%%%%%%
% Appendix
% %%%%%%%%%%%%%%%%%%%%%%%%%%%%%%%%%%%%%%%%%%%%%%%%%%%%%%%%%%
% %%%%%%%%%%%%%%%%%%%%%%%%%%%%%%%%%%%%%%%%%%%%%%%%%%%%%%%%%%

\appendix

\renewcommand{\thesection}{\Alph{section}.\arabic{section}}
\setcounter{section}{0}
\setcounter{table}{0}
\renewcommand{\thetable}{A\arabic{table}}
\renewcommand{\thefigure}{A\arabic{figure}}

\section{Online Appendix}

In this Appendix I include data details, other specifications, and robustness checks.

\begin{figure}[H]
    \centering
        \includegraphics[scale=0.75]{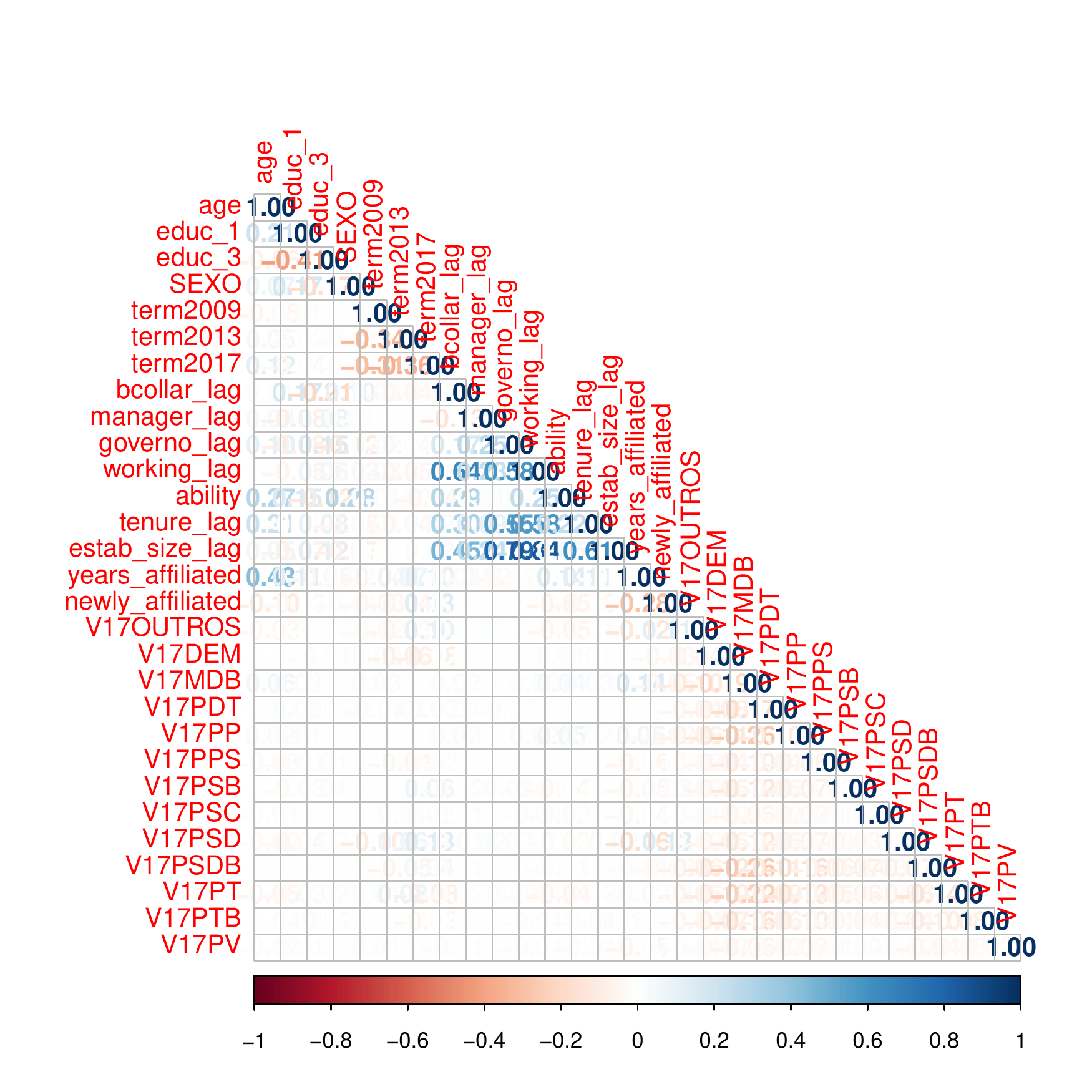}
    \caption{Correlation plot (benchmark specification)}
    \label{fig:corrplot}
\end{figure}

In Figure \ref{fig:corrplot}, I present the correlation plot for features in our benchmark specification. It is well known that machine learning procedures might work suboptimally when features are highly correlated among themselves. We see in Figure \ref{fig:corrplot} that this is not the case, and most features are close to orthogonal to each other, with the exception being characteristics of the previous employment. The only highly correlated features are establishment size of previous employment, whether previously employed and whether previously employed in the municipality.

\begin{figure}[H]
    \centering
        \includegraphics[scale=0.6]{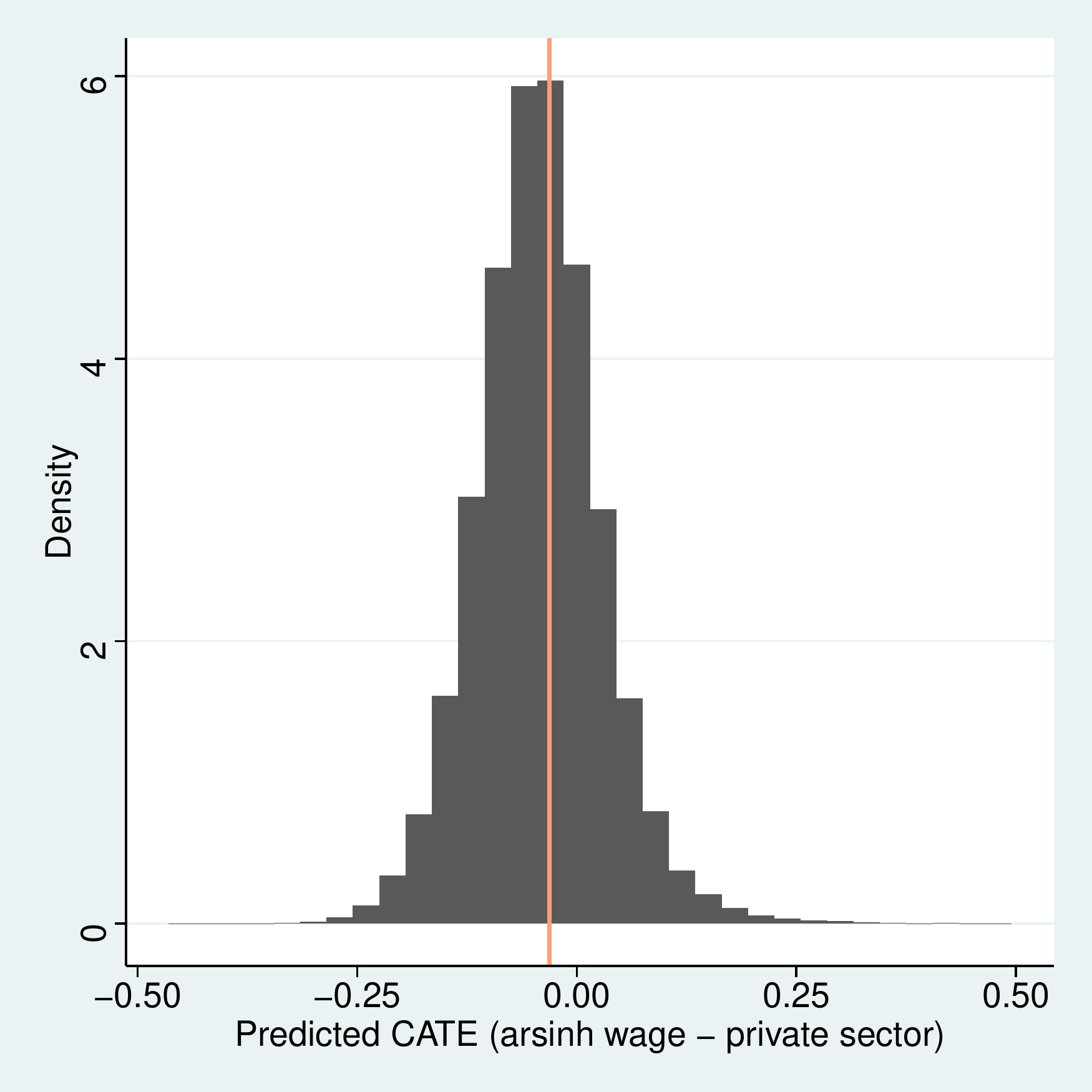}
    \caption{Predicted conditional average treatment effect for working in private sector with previous job in private sector}
    \label{fig:private}
\end{figure}

Figure \ref{fig:private} shows the histogram of conditional treatment effects of being affiliated to the mayor's party on working in the \textit{private} sector, for workers previously employed in the private sector. We see no positive effect.

Tables \ref{table:main_025} and \ref{table:main_001} show the same results as Table \ref{table:main_full} (the main table), but using as vote margin intervals $\pm 2.5$ percentage points and $\pm 1.0$ p.p., respectively. Encouragingly, we see similar results as in the benchmark analysis ($\pm 5$ p.p.).

Table \ref{tab:incumbent} shows the main results for a specification including incumbents, as in the benchmark analysis we include only first-year mayors.

Table \ref{table:ability_prev} is a robustness exercise on our measure of ability, estimating ability using only observations from before the election. (See Section \ref{sec:data} for a discussion of that variable.)

Finally, to complement Table \ref{table:private}, in the main text, which restricts the sample to workers previously employed in the private sector, we present here in Tables \ref{table:main_nowork} and \ref{table:main_public} the conditional treatment effects for workers previously unemployed and previously working in the municipal public sector, respectively.

%%%%%% TABLES

% Table created by stargazer v.5.2.2 by Marek Hlavac, Harvard University. E-mail: hlavac at fas.harvard.edu
% Date and time: Sun, Oct 31, 2021 - 09:34:08 AM
\begin{table}[h] \centering \footnotesize
\begin{threeparttable}
  \caption{Best linear predictor of conditional treatment effect estimates -- closer elections (2.5 p.p.)} 
  \label{table:main_025} 
\begin{tabular}{@{\extracolsep{5pt}}lccccc} 
\\[-1.8ex]\hline 
\hline \\[-1.8ex] 
 & \multicolumn{5}{c}{\textit{Dependent variable:}} \\ 
\cline{2-6} 
\\[-1.8ex] & (1) & (2) & (3) & (4) & (5)\\ 
\\[-1.8ex] & Municipal & Wage & Mun. empl.:  & Mun. empl.:  & Mun. empl.: \\ 
\\[-1.8ex] & employment &  (arsinh) & blue collar & white collar & manager \\ 
\hline \\[-1.8ex] 
Constant (ATE)& 0.088$^{}$ & 0.367$^{}$ & 0.010$^{}$ & 0.036$^{}$ & 0.043$^{}$ \\
  & (0.003) & (0.012) & (0.001) & (0.002) & (0.002) \\ 
  & & & & & \\  
Age & 0.026$^{}$ & 0.128$^{}$ & 0.003$^{}$ & 0.007$^{}$ & 0.016$^{}$ \\ 
  & (0.002) & (0.010) & (0.001) & (0.001) & (0.001) \\ 
  & & & & & \\ 
High-school & $-$0.067$^{}$ & $-$0.313$^{}$ & 0.010$^{}$ & $-$0.025$^{}$ & $-$0.051$^{}$ \\ 
incomplete  & (0.006) & (0.032) & (0.004) & (0.004) & (0.003) \\ 
  & & & & & \\  
University & 0.071$^{}$ & 0.449$^{}$ & $-$0.011$^{}$ & 0.008 & 0.078$^{}$ \\ 
  & (0.011) & (0.060) & (0.005) & (0.009) & (0.007) \\ 
  & & & & & \\ 
Male & 0.051$^{}$ & 0.250$^{}$ & 0.004 & 0.018$^{}$ & 0.028$^{}$ \\ 
  & (0.007) & (0.037) & (0.004) & (0.005) & (0.003) \\ 
  & & & & & \\ 
Ability & $-$0.004$^{}$ & 0.003 & $-$0.005$^{}$ & $-$0.001 & 0.001 \\ 
  & (0.002) & (0.010) & (0.001) & (0.001) & (0.001) \\ 
  & & & & & \\ 
Years since & $-$0.040$^{}$ & $-$0.154$^{}$ & $-$0.005$^{}$ & $-$0.017$^{}$ & $-$0.018$^{}$ \\   affiliation  & (0.002) & (0.011) & (0.001) & (0.002) & (0.001) \\ 
  & & & & & \\ 
  Newly affiliated & 0.008 & $-$0.010 & 0.016 & $-$0.008 & $-$0.006 \\ 
  & (0.037) & (0.188) & (0.017) & (0.025) & (0.020) \\ 
  & & & & & \\ 
  \multicolumn{6}{@{\extracolsep{5pt}}l}{\textit{Previous employment characteristics:}} \\ 
   & & & & & \\ 
Blue collar & $-$0.067$^{}$ & $-$0.228$^{}$ & 0.028$^{}$ & $-$0.085$^{}$ & $-$0.002 \\ 
  & (0.011) & (0.056) & (0.012) & (0.013) & (0.005) \\ 
  & & & & & \\ 
Manager & 0.540$^{}$ & 2.858$^{}$ & $-$0.007 & $-$0.228$^{}$ & 0.781$^{}$ \\ 
  & (0.053) & (0.257) & (0.021) & (0.037) & (0.066) \\ 
  & & & & & \\ 
Government & 0.139$^{}$ & 0.795$^{}$ & 0.017 & 0.071$^{}$ & 0.052$^{}$ \\ 
  & (0.017) & (0.105) & (0.012) & (0.011) & (0.008) \\ 
  & & & & & \\
Employed & $-$0.052$^{}$ & $-$0.424$^{}$ & $-$0.018$^{}$ & 0.034$^{}$ & $-$0.073$^{}$ \\ 
  & (0.014) & (0.082) & (0.010) & (0.011) & (0.010) \\ 
  & & & & & \\ 
Tenure & $-$0.038$^{}$ & $-$0.160$^{}$ & $-$0.004$^{}$ & $-$0.021$^{}$ & $-$0.013$^{}$ \\ 
  & (0.003) & (0.012) & (0.002) & (0.002) & (0.001) \\ 
  & & & & & \\ 
Establishment & 0.010$^{}$ & 0.083$^{}$ & $-$0.0001 & 0.005$^{}$ & 0.005 \\ 
size  & (0.004) & (0.028) & (0.002) & (0.003) & (0.003) \\ 
  & & & & & \\ 
\hline \\[-1.8ex] 
%Mean debiased error & 0.101 & 3.296 & 0.032 & 0.050 & 0.033 \\ 
%MDE / treated mean & 56.8\% & 264\% & 82.4\% & 64.5\% & 54.2\% \\ 
Mean forest & 0.943 & 0.957 & 0.821 & 0.954 & 0.948 \\ 
prediction  & (0.029) & (0.031) & (0.118) & (0.048) & (0.033) \\ 
Differential forest & 1.287 & 1.339 & 1.391 & 1.511 & 1.133 \\ 
prediction  & (0.035) & (0.036) & (0.201) & (0.071) & (0.064) \\ 
N & 330,048 & 330,048 & 330,048 & 330,048 & 330,048 \\
  & & & & & \\ 
\hline 
\hline \\[-1.8ex] 
\end{tabular} 
\begin{tablenotes}
\item \emph{Notes:} All regressions include municipality and term fixed effects, as well as party dummies. Considers only municipal elections with vote margin within $\left[-2.5\%, +2.5\%\right]$. The dependent variables capture the employment situation in December of the first year at office. This regression finds the best linear predictor to the conditional average treatment effect estimates of the random forest model, weighting the OLS with the estimated propensity scores. Since covariates are demeaned, the regression constants identify the average treatment effects. Coefficients (and standard errors) of dummy variables are then adjusted for a binary variable interpretation. Errors are clustered at the municipality level.
\end{tablenotes}
\end{threeparttable}
\end{table} 

% Table created by stargazer v.5.2.2 by Marek Hlavac, Harvard University. E-mail: hlavac at fas.harvard.edu
% Date and time: Sun, Oct 31, 2021 - 09:34:08 AM
\begin{table}[h] \centering \footnotesize
\begin{threeparttable}
  \caption{Best linear predictor of conditional treatment effect estimates -- closer elections (1 p.p.)}
  \label{table:main_001} 
\begin{tabular}{@{\extracolsep{5pt}}lccccc} 
\\[-1.8ex]\hline 
\hline \\[-1.8ex] 
 & \multicolumn{5}{c}{\textit{Dependent variable:}} \\ 
\cline{2-6} 
\\[-1.8ex] & (1) & (2) & (3) & (4) & (5)\\ 
\\[-1.8ex] & Municipal & Wage & Mun. empl.:  & Mun. empl.:  & Mun. empl.: \\ 
\\[-1.8ex] & employment &  (arsinh) & blue collar & white collar & manager \\ 
\hline \\[-1.8ex] 
Constant (ATE) & 0.090$^{}$ & 0.358$^{}$ & 0.012$^{}$ & 0.035$^{}$ & 0.044$^{}$ \\ 
  & (0.004) & (0.020) & (0.002) & (0.003) & (0.002) \\ 
  & & & & & \\  
Age & 0.022$^{}$ & 0.125$^{}$ & 0.001 & 0.007$^{}$ & 0.014$^{}$ \\ 
  & (0.003) & (0.015) & (0.002) & (0.002) & (0.001) \\ 
  & & & & & \\ 
High-school & $-$0.052$^{}$ & $-$0.260$^{}$ & 0.020$^{}$ & $-$0.022$^{}$ & $-$0.049$^{}$ \\ 
incomplete  & (0.009) & (0.051) & (0.007) & (0.006) & (0.005) \\ 
  & & & & & \\ 
University & 0.065$^{}$ & 0.377$^{}$ & $-$0.007 & $-$0.006 & 0.080$^{}$ \\ 
  & (0.019) & (0.099) & (0.009) & (0.015) & (0.011) \\ 
  & & & & & \\ 
Male & 0.062$^{}$ & 0.340$^{}$ & 0.009 & 0.018$^{}$ & 0.031$^{}$ \\ 
  & (0.011) & (0.059) & (0.007) & (0.008) & (0.005) \\ 
  & & & & & \\ 
 Ability & $-$0.005$^{}$ & $-$0.027 & $-$0.004$^{}$ & $-$0.003$^{}$ & 0.002 \\ 
  & (0.003) & (0.018) & (0.002) & (0.002) & (0.001) \\ 
  & & & & & \\
Years since & $-$0.040$^{}$ & $-$0.153$^{}$ & $-$0.006$^{}$ & $-$0.017$^{}$ & $-$0.017$^{}$ \\  affiliation  & (0.003) & (0.018) & (0.002) & (0.002) & (0.002) \\ 
  & & & & & \\ 
Newly affiliated & 0.025 & 0.121 & 0.037 & $-$0.007 & $-$0.010 \\ 
  & (0.056) & (0.277) & (0.028) & (0.037) & (0.026) \\ 
  & & & & & \\ 
\multicolumn{6}{@{\extracolsep{5pt}}l}{\textit{Previous employment characteristics:}} \\ 
   & & & & & \\ 
Blue collar & $-$0.082$^{}$ & $-$0.207$^{}$ & 0.020 & $-$0.082$^{}$ & $-$0.006 \\   & (0.016) & (0.096) & (0.016) & (0.018) & (0.008) \\ 
  & & & & & \\ 
Manager & 0.623$^{}$ & 3.561$^{}$ & $-$0.028 & $-$0.189$^{}$ & 0.873$^{}$ \\ 
  & (0.076) & (0.417) & (0.027) & (0.059) & (0.076) \\ 
  & & & & & \\  
Government & 0.143$^{}$ & 0.574$^{}$ & 0.022 & 0.061$^{}$ & 0.060$^{}$ \\ 
  & (0.028) & (0.167) & (0.019) & (0.016) & (0.012) \\ 
  & & & & & \\  
Employed & $-$0.040$^{}$ & $-$0.441$^{}$ & $-$0.015 & 0.047$^{}$ & $-$0.078$^{}$ \\ 
  & (0.020) & (0.121) & (0.013) & (0.016) & (0.011) \\ 
  & & & & & \\ 
Tenure & $-$0.036$^{}$ & $-$0.145$^{}$ & $-$0.005$^{}$ & $-$0.018$^{}$ & $-$0.013$^{}$ \\ 
  & (0.004) & (0.019) & (0.002) & (0.003) & (0.002) \\ 
  & & & & & \\ 
Establishment & 0.010 & 0.116$^{}$ & 0.002 & 0.0004 & 0.008$^{}$ \\ 
size  & (0.006) & (0.038) & (0.003) & (0.004) & (0.003) \\ 
  & & & & & \\ 
\hline \\[-1.8ex] 
%Mean debiased error & 0.101 & 3.342 & 0.033 & 0.049 & 0.031 \\ 
%MDE / treated mean & 56.9\% & 268\% & 81.2\% & 65.4\% & 53.9\% \\ 
Mean forest & 0.941 & 0.938 & 0.839 & 0.916 & 0.953 \\ 
prediction  & (0.043) & (0.052) & (0.148) & (0.072) & (0.045) \\ 
Differential forest & 1.413 & 1.549 & 1.313 & 1.493 & 1.212 \\ 
prediction  & (0.054) & (0.066) & (0.240) & (0.141) & (0.070) \\ 
N  & 129,164 & 129,164 & 129,164 & 129,164 & 129,164 \\ 
  & & & & & \\ 
\hline 
\hline \\[-1.8ex] 
\end{tabular} 
\begin{tablenotes}
\item \emph{Notes:} All regressions include municipality and term fixed effects, as well as party dummies. Considers only municipal elections with vote margin within $\left[-1.0\%, +1.0\%\right]$. The dependent variables capture the employment situation in December of the first year at office. This regression finds the best linear predictor to the conditional average treatment effect estimates of the random forest model, weighting the OLS with the estimated propensity scores. Since covariates are demeaned, the regression constants identify the average treatment effects. Coefficients (and standard errors) of dummy variables are then adjusted for a binary variable interpretation. Errors are clustered at the municipality level.
\end{tablenotes}
\end{threeparttable}
\end{table} 

% Table created by stargazer v.5.2.2 by Marek Hlavac, Harvard University. E-mail: hlavac at fas.harvard.edu
% Date and time: Sun, Oct 31, 2021 - 09:34:08 AM
\begin{table}[h] \centering \footnotesize
\begin{threeparttable}
  \caption{Best linear predictor of conditional treatment effect estimates -- data with incumbents} 
  \label{tab:incumbent} 
\begin{tabular}{@{\extracolsep{5pt}}lccccc} 
\\[-1.8ex]\hline 
\hline \\[-1.8ex] 
 & \multicolumn{5}{c}{\textit{Dependent variable:}} \\ 
\cline{2-6} 
\\[-1.8ex] & (1) & (2) & (3) & (4) & (5)\\ 
\\[-1.8ex] & Municipal & Wage & Mun. empl.:  & Mun. empl.:  & Mun. empl.: \\ 
\\[-1.8ex] & employment &  (arsinh) & blue collar & white collar & manager \\ 
\hline \\[-1.8ex] 
Constant (ATE) & 0.082$^{}$ & 0.345$^{}$ & 0.010$^{}$ & 0.034$^{}$ & 0.039$^{}$ \\ 
  & (0.001) & (0.007) & (0.001) & (0.001) & (0.001) \\ 
  & & & & & \\ 
Age & 0.023$^{}$ & 0.109$^{}$ & 0.003$^{}$ & 0.007$^{}$ & 0.014$^{}$ \\ 
  & (0.001) & (0.006) & (0.001) & (0.001) & (0.001) \\ 
  & & & & & \\ 
High-school & $-$0.061$^{}$ & $-$0.288$^{}$ & 0.008$^{}$ & $-$0.025$^{}$ & $-$0.044$^{}$ \\ 
incomplete  & (0.003) & (0.019) & (0.003) & (0.003) & (0.002) \\ 
  & & & & & \\ 
University & 0.062$^{}$ & 0.414$^{}$ & $-$0.013$^{}$ & 0.012$^{}$ & 0.063$^{}$ \\ 
  & (0.007) & (0.036) & (0.003) & (0.006) & (0.004) \\ 
  & & & & & \\
Male & 0.045$^{}$ & 0.187$^{}$ & 0.005$^{}$ & 0.015$^{}$ & 0.026$^{}$ \\ 
  & (0.004) & (0.022) & (0.002) & (0.003) & (0.002) \\ 
  & & & & & \\ 
Ability & 0.001 & 0.020$^{}$ & $-$0.003$^{}$ & 0.001 & 0.003$^{}$ \\ 
  & (0.001) & (0.006) & (0.001) & (0.001) & (0.001) \\ 
  & & & & & \\ 
Years since & $-$0.036$^{}$ & $-$0.141$^{}$ & $-$0.006$^{}$ & $-$0.016$^{}$ & $-$0.015$^{}$ \\  affiliation   & (0.001) & (0.007) & (0.001) & (0.001) & (0.001) \\ 
  & & & & & \\ 
Newly affiliated & 0.027 & 0.011 & 0.021$^{}$ & 0.005 & $-$0.003 \\ 
  & (0.022) & (0.117) & (0.011) & (0.015) & (0.012) \\ 
  & & & & & \\ 
\multicolumn{6}{@{\extracolsep{5pt}}l}{\textit{Previous employment characteristics:}} \\ 
   & & & & & \\ 
Blue collar & $-$0.082$^{}$ & $-$0.308$^{}$ & 0.029$^{}$ & $-$0.110$^{}$ & $-$0.001 \\ 
  & (0.007) & (0.039) & (0.007) & (0.008) & (0.003) \\ 
  & & & & & \\ 
Manager & 0.671$^{}$ & 2.758$^{}$ & $-$0.029$^{}$ & $-$0.326$^{}$ & 1.035$^{}$ \\ 
  & (0.031) & (0.150) & (0.014) & (0.024) & (0.035) \\ 
  & & & & & \\ 
Government & 0.165$^{}$ & 0.867$^{}$ & 0.022$^{}$ & 0.076$^{}$ & 0.069$^{}$ \\ 
  & (0.010) & (0.062) & (0.007) & (0.006) & (0.005) \\ 
  & & & & & \\
Employed & $-$0.036$^{}$ & $-$0.305$^{}$ & $-$0.022$^{}$ & 0.055$^{}$ & $-$0.072$^{}$ \\ 
  & (0.008) & (0.046) & (0.005) & (0.007) & (0.005) \\ 
  & & & & & \\ 
Tenure & $-$0.045$^{}$ & $-$0.171$^{}$ & $-$0.005$^{}$ & $-$0.025$^{}$ & $-$0.015$^{}$ \\ 
  & (0.002) & (0.008) & (0.001) & (0.001) & (0.001) \\ 
  & & & & & \\ 
Establishment & 0.017$^{}$ & 0.089$^{}$ & 0.001 & 0.008$^{}$ & 0.008$^{}$ \\ 
size  & (0.002) & (0.016) & (0.001) & (0.002) & (0.001) \\ 
  & & & & & \\ 
\hline \\[-1.8ex] 
Mean forest & 0.967 & 0.971 & 0.915 & 0.980 & 0.961 \\ 
prediction  & (0.017) & (0.019) & (0.072) & (0.030) & (0.020) \\ 
Differential forest & 1.136 & 1.126 & 1.427 & 1.263 & 1.104 \\ 
prediction  & (0.016) & (0.018) & (0.098) & (0.036) & (0.024) \\ 
N  & 907,796 &  907,796 & 907,796 & 907,796 & 907,796 \\ 
  & & & & & \\ 
\hline 
\hline \\[-1.8ex] 
\end{tabular} 
\begin{tablenotes}
\item \emph{Notes:} All regressions include municipality and term fixed effects, as well as party dummies. Considers only municipal elections with vote margin within $\left[-5.0\%, +5.0\%\right]$. The dependent variables capture the employment situation in December of the first year at office. This regression finds the best linear predictor to the conditional average treatment effect estimates of the random forest model, weighting the OLS with the estimated propensity scores. Since covariates are demeaned, the regression constants identify the average treatment effects. Coefficients (and standard errors) of dummy variables are then adjusted for a binary variable interpretation. Errors are clustered at the municipality level.
\end{tablenotes}
\end{threeparttable}
\end{table} 

% Table created by stargazer v.5.2.2 by Marek Hlavac, Harvard University. E-mail: hlavac at fas.harvard.edu
% Date and time: Sun, Oct 31, 2021 - 09:34:08 AM
\begin{table}[h] \centering \footnotesize
\begin{threeparttable}
  \caption{Best linear predictor of conditional treatment effect estimates -- ability calculated using only previous observations}
  \label{table:ability_prev} 
\begin{tabular}{@{\extracolsep{5pt}}lccccc} 
\\[-1.8ex]\hline 
\hline \\[-1.8ex] 
 & \multicolumn{5}{c}{\textit{Dependent variable:}} \\ 
\cline{2-6} 
\\[-1.8ex] & (1) & (2) & (3) & (4) & (5)\\ 
\\[-1.8ex] & Municipal & Wage & Mun. empl.:  & Mun. empl.:  & Mun. empl.: \\ 
\\[-1.8ex] & employment &  (arsinh) & blue collar & white collar & manager \\ 
\hline \\[-1.8ex] 
Constant (ATE) & 0.082$^{}$ & 0.349$^{}$ & 0.008$^{}$ & 0.033$^{}$ & 0.042$^{}$ \\ 
  & (0.002) & (0.010) & (0.001) & (0.002) & (0.001) \\ 
  & & & & & \\ 
Age & 0.020$^{}$ & 0.115$^{}$ & 0.001 & 0.005$^{}$ & 0.015$^{}$ \\ 
  & (0.001) & (0.007) & (0.001) & (0.001) & (0.001) \\ 
  & & & & & \\ 
High-school & $-$0.061$^{}$ & $-$0.284$^{}$ & 0.009$^{}$ & $-$0.024$^{}$ & $-$0.046$^{}$ \\ 
incomplete  & (0.005) & (0.028) & (0.003) & (0.003) & (0.003) \\ 
  & & & & & \\ 
University & 0.067$^{}$ & 0.492$^{}$ & $-$0.010$^{}$ & 0.012 & 0.069$^{}$ \\ 
  & (0.009) & (0.050) & (0.004) & (0.008) & (0.006) \\ 
  & & & & & \\ 
Male & 0.055$^{}$ & 0.262$^{}$ & 0.002 & 0.020$^{}$ & 0.031$^{}$ \\ 
  & (0.006) & (0.030) & (0.003) & (0.004) & (0.003) \\ 
  & & & & & \\
Previous ability & $-$0.024$^{}$ & $-$0.046 & $-$0.012$^{}$ & $-$0.006 & $-$0.007$^{}$ \\ 
  & (0.005) & (0.034) & (0.003) & (0.004) & (0.003) \\ 
  & & & & & \\ 
Years since & $-$0.039$^{}$ & $-$0.155$^{}$ & $-$0.004$^{}$ & $-$0.016$^{}$ & $-$0.019$^{}$ \\ 
affiliation  & (0.002) & (0.010) & (0.001) & (0.001) & (0.001) \\ 
  & & & & & \\ 
Newly affiliated & 0.003$^{}$ & 0.004 & 0.001$^{}$ & 0.001 & 0.001 \\ 
  & (0.002) & (0.009) & (0.001) & (0.001) & (0.001) \\ 
  & & & & & \\ 
\multicolumn{6}{@{\extracolsep{5pt}}l}{\textit{Previous employment characteristics:}} \\ 
   & & & & & \\ 
Blue collar & $-$0.015$^{}$ & $-$0.054$^{}$ & 0.005$^{}$ & $-$0.017$^{}$ & $-$0.001 \\
  & (0.002) & (0.010) & (0.002) & (0.002) & (0.001) \\ 
  & & & & & \\ 
Manager & 0.113$^{}$ & 0.550$^{}$ & $-$0.003 & $-$0.056$^{}$ & 0.173$^{}$ \\ 
  & (0.010) & (0.047) & (0.003) & (0.007) & (0.012) \\ 
  & & & & & \\ 
Government & 0.546$^{}$ & 3.551$^{}$ & 0.014 & 0.271$^{}$ & 0.261$^{}$ \\ 
  & (0.057) & (0.376) & (0.041) & (0.037) & (0.029) \\ 
  & & & & & \\ 
Employed & $-$0.029$^{}$ & $-$0.396$^{}$ & $-$0.029$^{}$ & 0.069$^{}$ & $-$0.074$^{}$ \\ 
  & (0.017) & (0.099) & (0.011) & (0.012) & (0.011) \\ 
  & & & & & \\ 
Tenure & $-$0.039$^{}$ & $-$0.165$^{}$ & $-$0.004$^{}$ & $-$0.023$^{}$ & $-$0.012$^{}$ \\ 
  & (0.002) & (0.012) & (0.001) & (0.002) & (0.001) \\ 
  & & & & & \\
Establishment & 0.010$^{}$ & 0.081$^{}$ & 0.003$^{}$ & 0.004$^{}$ & 0.002 \\ 
  & (0.003) & (0.022) & (0.002) & (0.002) & (0.002) \\ 
  & & & & & \\
\hline \\[-1.8ex] 
%Mean debiased error & 0.101 & 3.342 & 0.033 & 0.049 & 0.031 \\ 
%MDE / treated mean & 56.9\% & 268\% & 81.2\% & 65.4\% & 53.9\% \\ 
Mean forest & 0.960 & 0.958 & 0.864 & 0.961 & 0.961 \\ 
prediction  & (0.024) & (0.027) & (0.111) & (0.044) & (0.024) \\ 
Differential forest & 1.230 & 1.549 & 1.478 & 1.420 & 1.154 \\ 
prediction  & (0.026) & (0.026) & (0.136) & (0.052) & (0.040) \\ 
N  & 466,064 & 466,064 & 466,064 & 466,064 & 466,064 \\ 
  & & & & & \\ 
\hline 
\hline \\[-1.8ex] 
\end{tabular} 
\begin{tablenotes}
\item \emph{Notes:} All regressions include municipality and term fixed effects, as well as party dummies. Considers only municipal elections with vote margin within $\left[-5.0\%, +5.0\%\right]$. The dependent variables capture the employment situation in December of the first year at office. This regression finds the best linear predictor to the conditional average treatment effect estimates of the random forest model, weighting the OLS with the estimated propensity scores. Since covariates are demeaned, the regression constants identify the average treatment effects. Coefficients (and standard errors) of dummy variables are then adjusted for a binary variable interpretation. Errors are clustered at the municipality level.
\end{tablenotes}
\end{threeparttable}
\end{table} 

\begin{table}[h] \centering \small
\begin{threeparttable}
  \caption{Best linear predictor of conditional treatment effect estimates (previously unemployed or informal work)}  
  \label{table:main_nowork} 
\begin{tabular}{@{\extracolsep{5pt}}lccccc} 
\\[-1.8ex]\hline 
\hline \\[-1.8ex] 
 & \multicolumn{5}{c}{\textit{Dependent variable:}} \\ 
\cline{2-6} 
\\[-1.8ex] & (1) & (2) & (3) & (4) & (5)\\ 
\\[-1.8ex] & Municipal & Wage & Mun. empl.:  & Mun. empl.:  & Mun. empl.: \\ 
\\[-1.8ex] & employment &  (arsinh) & blue collar & white collar & manager \\ 
\hline \\[-1.8ex] 
Constant (ATE) & 0.095$^{}$ & 0.382$^{}$ & 0.011$^{}$ & 0.038$^{}$ & 0.046$^{}$ \\ 
  & (0.003) & (0.012) & (0.001) & (0.002) & (0.001) \\ 
  & & & & & \\ 
Age & 0.028$^{}$ & 0.131$^{}$ & 0.0002 & 0.006$^{}$ & 0.022$^{}$ \\ 
  & (0.002) & (0.009) & (0.001) & (0.001) & (0.001) \\ 
  & & & & & \\ 
High-school & $-$0.092$^{}$ & $-$0.441$^{}$ & 0.013$^{}$ & $-$0.042$^{}$ & $-$0.062$^{}$ \\ 
incomplete  & (0.006) & (0.029) & (0.004) & (0.004) & (0.003) \\ 
  & & & & & \\ 
University & 0.141$^{}$ & 0.876$^{}$ & $-$0.017$^{}$ & 0.044$^{}$ & 0.118$^{}$ \\ 
  & (0.012) & (0.068) & (0.005) & (0.010) & (0.008) \\ 
  & & & & & \\ 
Male & 0.052$^{}$ & 0.232$^{}$ & 0.006 & 0.005 & 0.042$^{}$ \\ 
  & (0.006) & (0.033) & (0.004) & (0.005) & (0.003) \\ 
  & & & & & \\ 
Years since & $-$0.044$^{}$ & $-$0.190$^{}$ & $-$0.005$^{}$ & $-$0.019$^{}$ & $-$0.021$^{}$ \\ 
affiliation  & (0.002) & (0.011) & (0.001) & (0.001) & (0.001) \\ 
  & & & & & \\ 
Newly affiliated & 0.050 & 0.171 & 0.049$^{}$ & 0.001 & $-$0.007 \\ 
  & (0.038) & (0.173) & (0.017) & (0.024) & (0.019) \\ 
  & & & & & \\ 
\hline \\[-1.8ex] 
%Mean debiased error & 0.099 & 3.280 & 0.032 & 0.049 & 0.031 \\ 
%MDE / treated mean & 57.6\% & 267\% & 82.7\% & 65.4\% & 54.6\% \\ 
Mean forest & 0.923 & 0.925 & 0.814 & 0.912 & 0.953 \\ 
prediction  & (0.028) & (0.030) & (0.089) & (0.041) & (0.026) \\ 
Differential forest & 1.206 & 1.221 & 1.120 & 1.106 & 1.197 \\ 
prediction  & (0.046) & (0.039) & (0.125) & (0.071) & (0.038) \\ 
N  & 327,347 & 327,347 & 327,347 & 327,347 & 327,347 \\ 
  & & & & & \\ 
\hline 
\hline \\[-1.8ex] 
\end{tabular} 
\begin{tablenotes}
\item \emph{Notes:} All regressions include municipality and term fixed effects, as well as party dummies. Considers only municipal elections with vote margin within $\left[-5\%, +5\%\right]$. Sample restricted to individuals who had no formal employment in the election year. The dependent variables are the values in the first year at office. This regression finds the best linear predictor to the conditional average treatment effect estimates of the random forest model, weighting the OLS with the estimated propensity scores.  Since covariates are demeaned, the regression constant identifies average treatment effects. Errors are clustered at the municipal level.
\end{tablenotes}
\end{threeparttable}
\end{table} 

% Table created by stargazer v.5.2.2 by Marek Hlavac, Harvard University. E-mail: hlavac at fas.harvard.edu
% Date and time: Fri, Oct 29, 2021 - 07:45:59 AM
\begin{table}[h] \centering 
\begin{threeparttable}
\small
\caption{Best linear predictor of conditional treatment effect estimates (previous job in public sector)} 
\label{table:main_public} 
\begin{tabular}{@{\extracolsep{5pt}}lccccc} 
\\ \hline 
\hline \\[-1.8ex] 
 & \multicolumn{5}{c}{\textit{Dependent variable:}} \\ 
\cline{2-6} 
\\[-1.8ex] & (1) & (2) & (3) & (4) & (5)\\ 
\\[-1.8ex] & Municipal & Wage & Mun. empl.:  & Mun. empl.:  & Mun. empl.: \\ 
\\[-1.8ex] & employment &  (arsinh) & blue collar & white collar & manager \\ 
\hline \\[-1.8ex] 
Constant (ATE) & 0.118$^{}$ & 0.597$^{}$ & 0.009$^{}$ & 0.047$^{}$ & 0.061$^{}$ \\ 
  & (0.004) & (0.017) & (0.003) & (0.003) & (0.003) \\ 
  & & & & & \\ 
Age & 0.027$^{}$ & 0.160$^{}$ & 0.008$^{}$ & 0.012$^{}$ & 0.008$^{}$ \\ 
  & (0.003) & (0.013) & (0.002) & (0.002) & (0.002) \\ 
  & & & & & \\ 
High-school & 0.001 & $-$0.091 & 0.017 & 0.019$^{}$ & $-$0.034$^{}$ \\ 
incomplete  & (0.013) & (0.055) & (0.012) & (0.010) & (0.007) \\ 
  & & & & & \\ 
University & $-$0.065$^{}$ & $-$0.063 & $-$0.003 & $-$0.067$^{}$ & 0.010 \\ 
  & (0.013) & (0.062) & (0.009) & (0.013) & (0.008) \\ 
  & & & & & \\ 
Male & 0.078$^{}$ & 0.388$^{}$ & 0.002 & 0.055$^{}$ & 0.022$^{}$ \\ 
  & (0.012) & (0.053) & (0.008) & (0.009) & (0.006) \\ 
  & & & & & \\ 
Ability & $-$0.010$^{}$ & $-$0.031$^{}$ & $-$0.009$^{}$ & $-$0.009$^{}$ & 0.006$^{}$ \\ 
  & (0.003) & (0.014) & (0.002) & (0.003) & (0.002) \\ 
  & & & & & \\ 
Years since & $-$0.028$^{}$ & $-$0.129$^{}$ & $-$0.004 & $-$0.013$^{}$ & $-$0.012$^{}$ \\ 
affiliation  & (0.003) & (0.015) & (0.002) & (0.003) & (0.002) \\ 
  & & & & & \\ 
Newly affiliated & 0.041 & $-$0.177 & 0.013 & 0.005 & 0.009 \\ 
  & (0.066) & (0.292) & (0.038) & (0.048) & (0.034) \\ 
  & & & & & \\ 
\multicolumn{6}{@{\extracolsep{5pt}}l}{\textit{Previous employment characteristics:}} \\ 
   & & & & & \\ 
Blue collar & $-$0.194$^{}$ & $-$0.776$^{}$ & 0.037$^{}$ & $-$0.172$^{}$ & $-$0.055$^{}$ \\ 
  & (0.014) & (0.058) & (0.016) & (0.016) & (0.007) \\ 
  & & & & & \\ 
Manager & 0.402$^{}$ & 1.939$^{}$ & $-$0.043$^{}$ & $-$0.265$^{}$ & 0.721$^{}$ \\ 
  & (0.030) & (0.139) & (0.015) & (0.026) & (0.033) \\ 
  & & & & & \\ 
Tenure & $-$0.073$^{}$ & $-$0.268$^{}$ & $-$0.013$^{}$ & $-$0.046$^{}$ & $-$0.014$^{}$ \\ 
  & (0.003) & (0.015) & (0.002) & (0.003) & (0.002) \\ 
  & & & & & \\ 
Establishment & 0.008 & 0.017 & 0.002 & 0.002 & 0.004 \\ 
size  & (0.005) & (0.020) & (0.003) & (0.004) & (0.003) \\ 
  & & & & & \\ 
\hline \\[-1.8ex] 
% Mean debiased error & 0.143 & 2.913 & 0.080 & 0.104 & 0.044 \\ 
% MDE / treated mean & 17.4\% & 75.16\% & 22.02\% & 28.6\% & 46.79\% \\ 
Mean forest & 1.008 & 1.018 & 0.895 & 1.042 & 0.986 \\ 
prediction  & (0.036) & (0.030) & (0.301) & (0.075) & (0.041) \\ 
Differential forest & 1.128 & 1.159 & 1.335 & 1.225 & 1.051 \\ 
prediction  & (0.029) & (0.032) & (0.189) & (0.052) & (0.042) \\ 
N  & 145,208 & 145,208 & 145,208 & 145,208 & 145,208 \\ 
  & & & & & \\ 
\hline 
\hline \\[-1.8ex] 
\end{tabular} 
\begin{tablenotes}
\item \emph{Notes:} All regressions include municipality and term fixed effects, as well as party dummies. Considers only municipal elections with vote margin within $\left[-5\%, +5\%\right]$. Sample restricted to individuals with private formal jobs in the previous year. The dependent variables are the values in the first year at office. This regression finds the best linear predictor to the conditional average treatment effect estimates of the random forest model, weighting the OLS with the estimated propensity scores. Since covariates are demeaned, the regression constant identifies average treatment effects. Errors are clustered at the municipal level.
\end{tablenotes}
\end{threeparttable}
\end{table} 

%%%%% FIGURES

% ==========================
% ==========================
% ==========================

\end{document}